\documentclass[11pt,a4paper]{article}

\usepackage{jcappub}
\usepackage{slashed}

\usepackage{multirow}
\usepackage{amssymb}
\usepackage{ulem}
\usepackage{cancel}
\usepackage{verbatim}
\usepackage{caption}
\usepackage{subcaption}
\usepackage[utf8]{inputenc}

\usepackage{array}
\newcolumntype{L}[1]{>{\raggedright\let\newline\\\arraybackslash\hspace{0pt}}m{#1}}
\newcolumntype{C}[1]{>{\centering\let\newline\\\arraybackslash\hspace{0pt}}m{#1}}
\newcolumntype{R}[1]{>{\raggedleft\let\newline\\\arraybackslash\hspace{0pt}}m{#1}}

\newcommand{\fett}[1]{\boldsymbol{#1}}

\newcommand{\dd}{{\rm{d}}}
\newcommand{\ii}{{\rm{i}}}

\newcommand{\be}{\begin{equation}}
\newcommand{\ee}{\end{equation}}
\newcommand{\e}{{\rm e}}
\newcommand{\LP}{{\rm P}}

\definecolor{darkred}{rgb}{0.5,0,0}
\definecolor{darkgreen}{rgb}{0,0.5,0}
\definecolor{darkblue}{rgb}{0,0,0.5}

\usepackage{hyperref}
\hypersetup{ colorlinks,
linkcolor=darkblue,
filecolor=darkgreen,
urlcolor=darkred,
citecolor=darkblue }

\newcommand{\inspire}[1]{\href{http://inspirehep.net/search?p=find+J+#1}
 {{\color{black}[{\color{blue} {\small in}SPIRE}]}}}
\newcommand{\book}[1]{\href{http://inspirehep.net/search?p=#1}
 {{\color{black}[{\color{blue} {\small in}SPIRE}]}}}

\newcommand{\inspired}[1]{\href{http://inspirehep.net/search?p=#1}
 {{\color{black}[{\color{blue} {\small in}SPIRE}]}}}

\newcommand{\CLASS}{\textsc{class}}

\begin{document}

%\begin{flushright}
%{\large \tt 
%TTK-12-15}
%\end{flushright}

\title{Interacting neutrinos in cosmology: exact description and constraints}

\date{\today}

\author[a]{Isabel M.~Oldengott,}
\emailAdd{ioldengott@physik.uni-bielefeld.de}

\author[b,c]{Thomas Tram,}
\emailAdd{thomas.tram@port.ac.uk}

\author[d,e]{Cornelius Rampf}
\emailAdd{rampf@thphys.uni-heidelberg.de}

\author[f]{and Yvonne Y.~Y.~Wong}
\emailAdd{yvonne.y.wong@unsw.edu.au}

\affiliation[a]{Fakult\"at f\"ur Physik, Bielefeld University, D--33501 Bielefeld, Germany}
\affiliation[b]{Department of Physics and Astronomy, University of Aarhus, Ny Munkegade 120, DK--8000 Aarhus C, Denmark}
\affiliation[c]{Institute of Cosmology and Gravitation, University of Portsmouth, Portsmouth PO1 3FX, United Kingdom}
\affiliation[d]{Institut f\"ur Theoretische Physik, Heidelberg University, Philosophenweg 16, D--69120 Heidelberg, Germany}
\affiliation[e]{Department of Physics, Israel Institute of Technology --- Technion, Haifa 32000, Israel}
\affiliation[f]{School of Physics, The University of New South Wales, Sydney NSW 2052, Australia}

\abstract{We consider the impact of neutrino self-interactions described by an effective four-fermion coupling on cosmological observations.
Implementing the exact Boltzmann hierarchy for interacting neutrinos first derived in~\cite{Oldengott:2014qra} into the Boltzmann solver \CLASS{}, 
we perform a detailed numerical analysis of the effects of the interaction on the cosmic microwave background (CMB) anisotropies, and
compare our results with known approximations in the literature.  While we find good agreement between our exact approach and the relaxation time approximation used in some recent studies, the popular $\left( c_{\text{eff}}^2,c_{\text{vis}}^2 \right)$-parameterisation fails to reproduce the correct scale dependence of the CMB temperature power spectrum.
We then proceed to derive constraints on the effective coupling constant $G_{\text{eff}}$ using currently available cosmological data via an MCMC analysis.  Interestingly, our results reveal a bimodal posterior distribution, where one mode represents the standard $\Lambda$CDM limit with $G_{\rm eff} \lesssim 10^8 \, G_{\rm F}$, and the other 
a  scenario in which neutrinos self-interact with an effective coupling constant $G_{\rm eff} \simeq 3 \times 10^9 \, G_{\rm F}$.}

\maketitle   

\flushbottom
%%%%%%%%%%%%%%%%%%%%%%%%%%%%%%%%%%%%%%%%%%%%%%%%%%%%%%%%%%

\section{Introduction}
\label{Introduction}

The nature of neutrinos and especially the mechanism by which they acquire  masses constitute some of the most long-standing puzzles in particle physics.
 Because of the assumption that only left-handed neutrinos exist, neutrinos have been incorporated into the standard model (SM) of particle physics as exactly massless particles. However, the discovery of neutrino oscillations has long since refuted the assumption of exact masslessness.  Besides providing a clear hint of physics beyond the SM, 
 this also suggests that any extension to the SM to account for neutrino masses necessitates new coupling of the neutrino to particles as yet unobserved. 
 
 Numerous models of neutrino mass generation have been proposed in the literature. One interesting direction are majoron-like models in which a neutrino mass term is generated by the spontaneous breaking of a $U(1)_{B-L}$ symmetry~\cite{Gelmini:1980re,Choi:1991aa,Acker:1992eh,Chikashige:1980ui,Georgi:1981pg}. The symmetry breaking is accompanied by the appearance of a new Goldstone boson, called the majoron, that primarily couples to neutrinos via the Yukawa interaction,
\begin{align}
\mathcal{L}_{\rm int} &= \mathfrak{g}_{ij} \bar{\nu}_i \nu_j \phi +\mathfrak{h}_{ij} \bar{\nu}_i \gamma_5 \nu_j \phi  \,, \label{Lagrangian}
\end{align}
where $\mathfrak{g}_{ij}$ and $\mathfrak{h}_{ij}$ are, respectively, the scalar and pseudo-scalar couplings.
Constraints on interactions of the type~\eqref{Lagrangian} have been derived from astrophysics (e.g., \cite{Ng:2014pca,Ioka:2014kca}), big bang nucleosynthesis (e.g., \cite{Aarssen:2012fx}), neutrinoless double $\beta$-decay (e.g., \cite{Gando:2012pj}), as well as the decay widths of the $Z$ boson and certain mesons (e.g., \cite{Abe:2013hdq,Laha:2013xua,Lessa:2007up}). 
Observations of the cosmic microwave background (CMB) temperature and polarisation anisotropies have so far provided useful insights into neutrino physics (e.g., limiting the sum of neutrino masses to $\sum m_{\nu} \lesssim 0.23$\,eV~\cite{Ade:2015xua}); it is therefore interesting to ponder if CMB measurements might also be sensitive to new neutrino interactions.

The generic signature of neutrino interactions during the time of CMB formation is an enhancement of its temperature power spectrum at multipoles $\ell \gtrsim 200$  following from a simple argument (see e.g.~\cite{Hannestad:2004qu,Bell:2005dr,Cyr-Racine:2013jua}). Non-interacting (and hence free-streaming) neutrinos in an inhomogeneous spacetime engender shear stress in the neutrino fluid, erasing fluctuations that might initially be present in the fluid and suppressing further growth.
In contrast, neutrino interactions tend to isotropise the neutrino fluid locally, allowing its energy density contrast and velocity divergence to 
undergo acoustic oscillations in the sub-horizon limit.  This in turn enhances the energy density and velocity contributions to the total gravitational source, and consequently the amplitude of the CMB temperature fluctuations on all scales that entered the horizon at times prior to photon decoupling.

While the limiting behaviours are well understood and their signatures easy to predict, the transition from fully interacting to non-interacting (or vice versa) is less clearcut.
Several previous works have attempted to place CMB constraints on new neutrino interactions using a variety of heuristic arguments to model the equations of motion for the neutrino perturbations, the so-called ``Boltzmann hierarchy'', in the transition region~\cite{Hannestad:2004qu,Bell:2005dr,Friedland:2007vv,Basboll:2008fx,Archidiacono:2011gq,Smith:2011es,Diamanti:2012tg,Cyr-Racine:2013jua}. Here, we take the view that once the interaction has been specified, the correct Boltzmann hierarchy should follow automatically from the collisional Boltzmann equation; the challenge lies in reducing the collisional integral to a numerically tractable form.

In a previous work~\cite{Oldengott:2014qra},  some of us  determined the Boltzmann hierarchy in the presence of new neutrino interactions of the type~\eqref{Lagrangian} to first order in the perturbed quantities and in two limits of the scalar particle mass---extremely massive and effectively massless, relative to the typical energies of the neutrinos. 
In the present work we shall focus on the former, the ``massive scalar'' limit, wherein the interaction between two neutrinos becomes effectively a four-fermion interaction.
We implement the corresponding neutrino Boltzmann hierarchy in the Boltzmann solver {\sc class}~\cite{Blas:2011rf}, and present a detailed numerical analysis of the signatures of interacting neutrinos in the CMB anisotropies.   We also take this opportunity to update the constraints on  neutrino interactions 
using the 2015 data from the Planck CMB mission~\cite{Ade:2015xua} and other recent cosmological observations.

The paper is organised as follows.  We begin in section~\ref{Formalism} with a review of the neutrino Boltzmann hierarchy first derived in~\cite{Oldengott:2014qra}, followed by a brief summary of other approaches in the literature.  In section \ref{Implementation} we describe the implementation of the hierarchy in the Boltzmann solver~{\sc class}, and present for the first time numerical calculations of the CMB anisotropies in the presence of neutrino self-interactions using this approach.  We derive constraints on the interaction strength from cosmological observations  in section~\ref{Constraints on the neutrino coupling}.  Section~\ref{Conclusions} contains our conclusions.

%%%%%%%%%%%%%%%%%%%%%%%%%%%%%%%%%%%%%%%%%%%%%%%%%%%%%%%%%

\section{Formalism} 
\label{Formalism}

\begin{figure}[t]
\centering
\includegraphics[width=\textwidth]{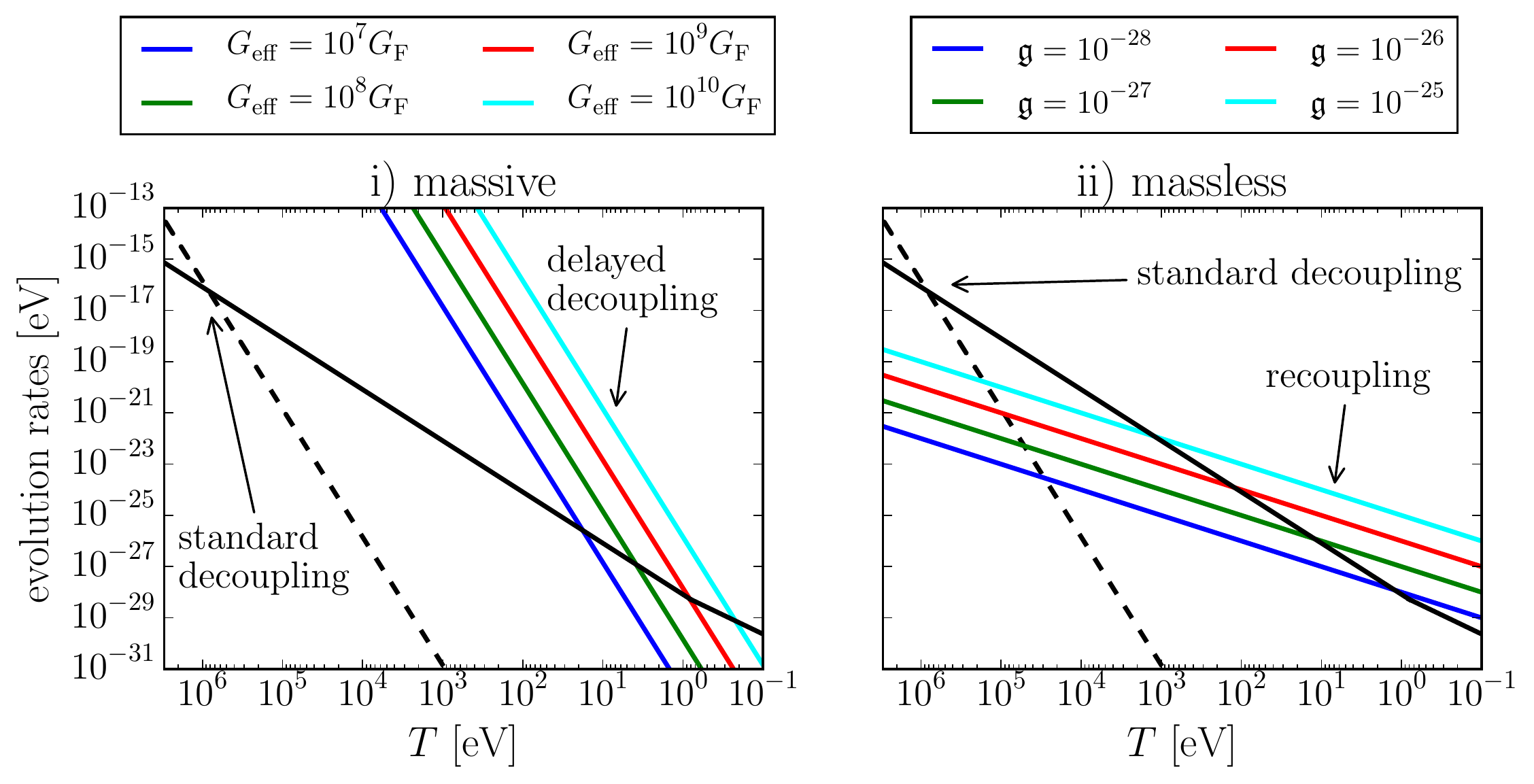}
\caption{Interaction rates per particle (coloured), in comparison with the Hubble expansion rate (black) and the standard weak interaction rate (dashed), assuming, in the left panel, a massive scalar particle  and, in the right, a massless scalar particle.}
\label{decoupling_plot}
\end{figure}

In a previous paper~\cite{Oldengott:2014qra}, some of us established the formal framework to study the impact of neutrino interactions on the CMB anisotropies in two limiting 
scenarios:
\begin{itemize}
\item[(i)] The scalar mass far exceeds the typical energies of the neutrinos in the CMB epoch. In this limit the Lagrangian \eqref{Lagrangian} becomes effectively a four-fermion interaction; any initial population of scalar particles will have decayed radiatively, and repopulation is kinematically suppressed. It  therefore suffices to consider only the neutrino self-interaction $\nu \nu \rightarrow \nu \nu$, while treating the $\phi$ population as essentially non-existent. In this scenario, neutrinos decouple from the rest of the cosmic plasma at the weak decoupling temperature, but remain scattering with each other at an interaction rate per particle given by $\Gamma_{\rm m} \sim  \mathfrak{g}^4 T^5_\nu/m_\phi^4 \equiv G_{\text{eff}}^2 T^5_\nu$, assuming $G_{\text{eff}} > G_{\rm F}$, where $G_{\rm F}$ is the Fermi constant.

\item[(ii)] An effectively massless scalar relative to the typical energies of the neutrinos.  With an interaction rate  per particle of $\Gamma_{\text{0}} \sim \mathfrak{g}^4 T_{\nu}$, the neutrinos in this scenario  decouple at the weak decoupling temperature, free-stream for some time, and then recouple when $\Gamma_{\text{0}}$ overtakes the Hubble expansion rate $H \sim T^2/m_{\text{Pl}}$, where $m_{\rm Pl}$ is the Planck mass. Because the production of scalar particles is now kinematically possible, this scenario is {\it a priori} numerically far less tractable than the massive scalar case~(i), and we shall not consider it in the present work.

\end{itemize}
The decoupling and, where applicable, recoupling behaviours of both limiting scenarios are illustrated in figure~\ref{decoupling_plot}.

%%%%%%%%%%%%%%%

\subsection{Neutrino Boltzmann hierarchy: Massive scalar limit}\label{sec:hierarchy}

Following the notation of~\cite{Ma:1995ey}, and working in the synchronous gauge defined by the line element
$\dd s^2 = a^2(\eta) \left[ - \dd \eta^2 + \left( \delta_{ij} + h_{ij}(\fett{x},\eta) \right) \right] \dd x^i \dd x^j$, 
the linear Boltzmann hierarchy for neutrinos interacting via the exchange of a very massive scalar particle is given by~\cite{Oldengott:2014qra}
\begin{equation}
\begin{aligned} 
%\allowdisplaybreaks
\label{finalBoltzmann_mass}
 \dot{\Psi}_{0}(q) =&  - k \Psi_{1}(q) + \frac{1}{6} \frac{\partial \ln \bar{f}}{\partial \ln q} \dot{h} - \frac{80}{3} \frac{\mathrm{N} T_{\nu,0}^5 G_{\text{eff}}^2}{a^4(2\pi)^{3}} \, q \,
 \Psi_{0}(q)  \\
 &\hspace{10mm}+ \frac{4\mathrm{N} T_{\nu,0}^5 G_{\text{eff}}^2}{a^4(2\pi)^{3}} \int  \dd q' \,  \left[ K^{\rm m}_0(q,q') - \frac{10}{9} q^2\, {q'}^2 \e^{-q}
  \right] \, \frac{q' \bar{f}(q')}{q \bar{f}(q)}  \, \Psi_{0}(q') \, , \\ 
\dot{\Psi}_{1}(q) =&  - \frac{2}{3}k \Psi_{2}(q)+ \frac{1}{3}k \Psi_{0}(q) - \frac{80}{3} \frac{\mathrm{N} T_{\nu,0}^5 G_{\text{eff}}^2}{a^4(2\pi)^{3}} \, q \, \Psi_{1}(q) \\
&\hspace{10mm}+ \frac{4\mathrm{N} T_{\nu,0}^5 G_{\text{eff}}^2}{a^4(2\pi)^{3}} \int  \dd q' \, \left[ K^{\rm m}_1(q,q')
+\frac{5}{9} q^2 \, {q'}^2 \e^{-q} \right] \, \frac{q' \bar{f}(q')}{q \bar{f}(q)}  \, \Psi_{1}(q') \, , \\ 
\dot{\Psi}_{2} (q) =&  - \frac{3}{5}k \Psi_{3}(q) + \frac{2}{5}k \Psi_{1}(q)- \frac{\partial \ln \bar{f}}{\partial \ln q} \left(\frac{2}{5} \dot{\tilde{\eta}}+ \frac{1}{15} \dot{h} \right) -  \frac{80}{3} \frac{\mathrm{N} T_{\nu,0}^5 G_{\text{eff}}^2}{a^4(2\pi)^{3}} \,q \, \Psi_{2}(q) \\
&\hspace{10mm} + \frac{4\mathrm{N} T_{\nu,0}^5 G_{\text{eff}}^2}{a^4(2\pi)^{3}} \int  \dd q' \,  \left[ K^{\rm m}_2(q,q') - \frac{1}{9} q^2 \, {q'}^2 \e^{-q}\right] \, \frac{q' \bar{f}(q')}{q \bar{f}(q)} \, \Psi_{2}(q') \, , \\
\dot{\Psi}_{\ell>2}(q) =& \frac{k}{2\ell+1} \left[ \ell \Psi_{\ell-1}(q)-(\ell+1)\Psi_{\ell+1}(q) \right] 
  - \frac{80}{3} \frac{\mathrm{N} T_{\nu,0}^5 G_{\text{eff}}^2}{a^4(2\pi)^{3}} \, q \, \Psi_{\ell}(q) \\
&\hspace{10mm} + \frac{4\mathrm{N} T_{\nu,0}^5 G_{\text{eff}}^2}{a^4(2\pi)^{3}} \int  \dd q' \, K^{\rm m}_\ell(q,q') \, \frac{q' \bar{f}(q')}{q \bar{f}(q)}  \, \Psi_{\ell}(q') \, .
\end{aligned}
\end{equation}
Here, an overdot denotes a derivative with respect to the conformal time $\eta$,   $h \equiv \delta^{ij} h_{ij}(\fett{k},\eta)$ and $\tilde \eta \equiv \tilde \eta(\fett{k},\eta) = -k^i k^j h_{ij}/(4k^2) + h/12$ are, respectively, the trace and traceless perturbation of $h_{ij}$ in Fourier space,
and $\Psi_\ell (k,q,\eta)$ is the $\ell$th Legendre moment, defined via
\be
 \Psi(q) \equiv \Psi(k,q,\cos \epsilon, \eta) 
  = \sum_{\ell =0}^\infty (-\ii)^\ell (2 \ell +1) \Psi_\ell(k,q,\eta) {\rm P}_\ell(\cos \epsilon) \,,
\ee
of the phase space density $f = \bar{f}(1+\Psi)$, with $\cos \epsilon = \fett{k} \cdot \fett{q}/(k q)$ and ${\rm P}_\ell(\cos \epsilon)$
a Legendre polynomial of order $\ell$.

%%%%%%
%%%%%%

In the collision terms, i.e., terms proportional to $G_{\rm eff}^2$, note that for notational simplicity and in contrast to~\cite{Oldengott:2014qra}, we have absorbed the present-day neutrino temperature $T_{\nu,0}=1.95$~K into the definition of the comoving momentum variable~$q \equiv |\fett{q}|$.  Accordingly, the integral kernels $K^{\rm m}_{\ell}(q,q')$ are now defined as
\begin{equation}
K^{\rm m}_{\ell}(q,q')= \int_{-1}^1 \dd \cos \theta \, K^{\rm m}(q,q',\cos \theta) \, \LP_{\ell}(\cos \theta), 
\label{Kernel_Legendre}
\end{equation}
where 
\begin{equation}
\begin{aligned}
K^{\rm m}(q, q', \cos \theta) &\equiv
 \frac{1}{16 P^5} \, \e^{-(Q_-+P)/2} \;  \left(Q_-^2-P^2 \right)^2  \\
 &\qquad \times \Big[ P^2 \left(3 P^2 - 2 P-4 \right)+ Q_+^2 \left( P^2 + 6 P + 12 \right)
 \Big]\,,
\end{aligned}
\end{equation}
with the variables $P \equiv |\fett{q}-\fett{q'}|$ and $Q_{\pm} \equiv q \pm q'$.   Furthermore, the normalisation factor $\mathrm{N}$, previously defined such that the background distribution~$\bar f$ matches the {\it number density} of a relativistic Fermi--Dirac distribution, is now defined to match the relativistic Fermi--Dirac {\it energy density}, i.e., 
\begin{equation}
\mathrm{N} \equiv \frac{7 \pi^4}{720}.
\end{equation}
We also acknowledge here that, independently of these new definitions, a factor of $2$ had been omitted in front of the integral terms by oversight in~\cite{Oldengott:2014qra}.

Note that the $\ell=0,1$ equations of the Boltzmann hierarchy~(\ref{finalBoltzmann_mass}) must satisfy the conservation of number, energy, and momentum, which is reflected in the complete cancellation of their respective collision terms when integrated over momentum in the appropriate manner. These requirements can be used as a formal check of the correctness of the integral kernels \eqref{Kernel_Legendre}. See appendix~\ref{Proof of number-, energy- and momentum conservation} for details.

Lastly, for completeness, we remind the reader that equation~(\ref{finalBoltzmann_mass}) has been derived under the assumption of (i) no Pauli blocking or Bose enhancement, (ii) Maxwell--Boltzmann distribution for non-degenerate background distribution functions, (iii) Majorana neutrinos that are ultra-relativistic in the timeframe of interest, and (iv) flavour-independent and diagonal scalar coupling, i.e., $\mathfrak{g}_{ij} \equiv \mathfrak{g}  \,\delta_{ij}$, and no pseudo-scalar coupling, i.e., $\mathfrak{h}_{ij}=0$. Modifying any one of these assumptions may  substantially alter the form of the hierarchy.

%%%%%%%%%%%%%%%%%%%%%%%%%%%%%%%%%%%

\subsection{Comparison to previous works}\label{sec:comparison}

The salient feature of our ``exact'' Boltzmann hierarchy \eqref{finalBoltzmann_mass} is that it has a momentum dependence arising from the non-negligible energy transfer that accompanies a neutrino-neutrino scattering event. Such a momentum dependence does not appear in the Thomson scattering limit,
%such as in the analogous hierarchy for the photon perturbations due to photon-electron scattering at linear order in perturbation theory\footnote{The Boltzmann hierarchy for photons is also momentum-dependent, however, at second order in perturbation theory~\cite{Beneke:2010eg} and beyond.}, 
and, to our knowledge, has been ignored in all earlier models of interacting neutrinos in cosmology. We briefly summarise these models below.  A more detailed discussion can be found in~\citep{Oldengott:2014qra}.

\paragraph{$\left( c_{\text{eff}}^2,c_{\text{vis}}^2 \right)$-parameterisation}
This most commonly-used model introduces two new parameters in the neutrino Boltzmann hierarchy, namely, an effective sound speed $c_{\text{eff}}$ and a viscosity parameter $c_{\text{vis}}$.  In the synchronous gauge, the modified hierarchy reads
 \begin{equation}
 \begin{aligned}
 \dot{\delta}_{\nu} &= -\frac{4}{3} \theta_\nu- \frac{2}{3}\dot{h} + \frac{\dot{a}}{a}(1-3 c_{\text{eff}}^2) \left( \delta_{\nu}+ 4 \frac{\dot{a}}{a} \frac{\theta_{\nu}}{k^2} \right) \ ,  \\ 
 \dot{\theta}_{\nu}&=k^2 \left(\frac{1}{4} \delta_\nu- \sigma_\nu \right) - \frac{k^2}{4} (1-3 c_{\rm eff}^2)  \left( \delta_{\nu}+ 4 \frac{\dot{a}}{a} \frac{\theta_{\nu}}{k^2} \right) \ , \\
  \dot{\cal F}_{\nu2} &= 2 \dot{\sigma}_\nu= \frac{8}{15} \theta_\nu  - \frac{3}{5} k {\cal F}_{\nu 3} +\frac{4}{15} \dot{h} + \frac{8}{5} \dot{\tilde \eta}- (1-3 c_{\rm vis}^2) \left(\frac{8}{15} \theta_\nu +\frac{4}{15} \dot{h} + \frac{8}{5} \dot{\tilde \eta}  \right) \ ,\\
    \dot{\cal F}_{\nu \ell} &= \frac{k}{2 \ell + 1} \left[ \ell {\cal F}_{\nu (\ell -1)} - (\ell + 1) {\cal F}_{\nu (\ell +1)} \right] \ , \quad \ell \geq 3 
 \end{aligned}
\label{ceff_cvis_parametrisation} 
\end{equation}
where we have used the notation of~\cite{Ma:1995ey}.
 This parameterisation was first introduced in~\cite{Hu:1998kj} to describe a generalised dark matter model, and reinterpreted in the context of neutrino interactions in, e.g.,~\cite{Archidiacono:2011gq,Smith:2011es,Diamanti:2012tg}.

Setting $c_{\text{eff}}^2=c_{\text{vis}}^2=1/3$ in equation~(\ref{ceff_cvis_parametrisation}) reproduces the limit of  free-streaming, non-interacting neutrinos, whereas the case of~$c_{\text{vis}}^2=0$ mimics the tightly-coupled limit---provided that the $\ell \geq 2$ multipoles are initially unpopulated. 
This last condition on the $\ell\geq 2$ multipoles should hold in the massive scalar scenario considered in this work, wherein neutrino decoupling is merely delayed by the new interaction. In the massless scalar scenario, however, the new interaction serves  to recouple neutrinos that have already been free-streaming for some time, which clearly violates the condition on $\ell \geq2$.  Furthermore,
 in the context of an isolated ultra-relativistic system of particles, self-interacting or otherwise, any choice for $c_{\rm eff}$ other than $c_{\text{eff}}^2=1/3$ would not comply with conservation of momentum, and is therefore not physically meaningful. 

Doubts about the physical meaningfulness of the $\left( c_{\text{eff}}^2,c_{\text{vis}}^2 \right)$-model  have been previously expressed in~\citep{Oldengott:2014qra,Cyr-Racine:2013jua,Sellentin:2014gaa}.  Indeed, we shall show in section~\ref{Impact of interacting neutrinos on the CMB} that not only does the  parameterisation~\eqref{ceff_cvis_parametrisation} have no formal interpretation in terms of particle scattering, it does not even reproduce the correct CMB phenomenology due to neutrino scattering.

\paragraph{Separable ansatz} 
In \cite{Cyr-Racine:2013jua} a damping term proportional to the rate of change of the neutrino opacity,
$\dot \tau_\nu \equiv -a G_{\rm eff}^2 T_\nu^5$, is introduced in the neutrino Boltzmann hierarchy at orders $\ell \geq 2$, i.e.,
\begin{equation}
\begin{aligned}
\label{eq:cyr}
\dot{\cal F}_{\nu 2} & =\frac{8}{15} \theta_\nu  - \frac{3}{5} k {\cal F}_{\nu 3} +\frac{4}{15} \dot{h} + \frac{8}{5} \dot{\tilde \eta}+ \alpha_2 \dot{\tau}_\nu {\cal F}_{\nu 2}\ , \\
\dot{\cal F}_{\nu \ell} &= \frac{k}{2 \ell + 1} \left[ \ell {\cal F}_{\nu (\ell -1)} - (\ell + 1) {\cal F}_{\nu (\ell +1)} \right] + \alpha_\ell \dot{\tau}_\nu {\cal F}_{\nu \ell}\ , \quad \ell \geq 3 \ ,
\end{aligned}
\end{equation}
where the $\alpha_{\ell}$s are model-dependent coefficients of order unity.  The monopole and dipole equations remain unaltered on account of energy and momentum conservation. 
While equation~(\ref{eq:cyr}) is clearly motivated by the first-order Boltzmann hierarchy for photons, we observe that it can be obtained from the exact Boltzmann hierarchy~(\ref{finalBoltzmann_mass}) by applying the ansatz that the phase space perturbation $\Psi(q)$ is independent of momentum, or, equivalently,
\begin{equation}
\label{eq:ansatz}
\Psi_{\ell}(k,q,t) \approx - \frac{1}{4} \frac{\mathrm{d} \ln \bar{f}}{\mathrm{d} \ln q} \mathcal{F}_{\ell}(k,t).
\end{equation}
Then, the two approaches are related simply by an integration in momentum.  We call this the ``separable ansatz''.  See also~\cite{Cyr-Racine:2013jua}.
 
The separable ansatz~(\ref{eq:ansatz}) also allows us to compare the exact Boltzmann hierarchy~\eqref{finalBoltzmann_mass} with equation~\eqref{eq:cyr} in a meaningful way, as it enables us to compute explicitly the model-dependent coefficients $\alpha_\ell$.  For a flavour-independent and diagonal interaction described by the scalar part of the Lagrangian~(\ref{Lagrangian}), we find
 \begin{equation}
\begin{aligned}
\alpha_{2} &= 0.40, \\
\alpha_{3} &= 0.43, \\
\alpha_{4} &= 0.46, \\
\alpha_{5} &= 0.47, \\
\alpha_{\ell \geq 6} &= 0.48.
\end{aligned}
\label{alphas}
\end{equation}
We shall use these coefficients when implementing equation~(\ref{eq:cyr}) in {\sc class} in section~\ref{Implementation}.

Note that equation~\eqref{eq:cyr} can also be motivated via the so-called relaxation time approximation (RTA, also sometimes referred to as the Bhatnagar--Gross--Krook approximation~\cite{Bhatnagar:1954zz}), which assumes that the perturbed collision integral  takes the form
\begin{equation}
{\cal C}[f] \approx -\frac{{\cal F}}{\tau_{\text{rel}}},
\label{RTA}
\end{equation}
where $\tau_{\text{rel}}$ is the time the system takes to relax to its equilibrium configuration.  In fact, we note that the momentum-dependent version of equation~(\ref{eq:cyr}), first presented in~\cite{Hannestad:2000gt} in the context of self-interacting warm dark matter, used this approximation. 
Assuming a momentum-independent $\tau_{\text{rel}}^{-1} = - \alpha_{\ell} \dot{\tau}_{\nu}$, it is easy to see that the RTA will lead to the Boltzmann hierarchy~\eqref{eq:cyr} upon integration in momentum.  
The RTA is often used in non-equilibrium physics  to understand the basic features of thermalisation processes.  It is, however, also known to be an  incomplete description of the dissipative dynamics of generic systems of particles~\cite{Bazow:2016oky}. 

In the following we shall refer to the model~(\ref{eq:cyr}) supplemented by the coefficients~(\ref{alphas}) as either the separable ansatz or the RTA.

%%%%%%%%%%%%%%%%%%%%%%%%%%%%%%%%%%%%%%%%%%%%%%%%%%%%%%%%%

\section{Implementation} 
\label{Implementation}

\subsection{Boltzmann hierarchy on a momentum grid}

We have implemented the exact Boltzmann hierarchy~\eqref{finalBoltzmann_mass} in the Boltzmann solver {\sc class} in order to study the impact of neutrino  self-interactions on the CMB anisotropies.  The implementation requires that we discretise the momentum variable $q$, and numerically solve the Boltzmann hierarchy for each momentum bin $q_i$.   We describe in this section how this can be achieved in a numerically stable fashion.

The contribution to the perturbed energy-momentum tensor is obtained by summation over all momentum bins, e.g., the perturbed energy density of the neutrinos is given by
\begin{equation}
\delta \rho = \int  \dd q \, q^3  \,\bar{f}(q) \, \Psi_{0}(q) \approx \sum_i q_i^3 w_i \Psi_{0,i},
\end{equation} 
where we have defined $\Psi_{\ell}(q_i) \equiv \Psi_{\ell,i}$.
 The integral weights $w_i$ depend on the integration method used, and incorporate by definition the background distribution function $\bar{f}_i \equiv \bar{f}(q_i)$ of the neutrinos. 
For the present numerical implementation we have chosen a uniform momentum grid and a trapezoidal integration method.
The discretised Boltzmann equation can then be written as
\begin{equation}
\dot{\Psi}_{\ell,i} = G_{\ell,i} + \frac{2\mathrm{N} T_{\nu,0}^5 G_{\text{eff}}^2}{a^4(2\pi)^{3}} \sum_j  M_{\ell,ij} \Psi_{\ell,j},  
\label{matrix_equation}
\end{equation} 
where we have defined
\begin{equation}
\begin{aligned}
G_{0,i} = & -k \Psi_{1,i}+ \frac{\partial \ln \bar{f}}{\partial \ln q_i} \dot{h}, \\
G_{2,i} = & \frac{2}{5} k \Psi_{1,i} -\frac{3}{5} k  \Psi_{3,i} + \frac{\partial \ln \bar{f}}{\partial \ln q_i} \left(\frac{2}{5} \dot{\tilde{\eta}}+ \frac{1}{15} \dot{h} \right), \\
G_{\ell,i} = & \frac{k}{2\ell+1} \left[ \ell \Psi_{\ell-1,i}-(\ell+1)\Psi_{\ell+1,i} \right]  \hspace{5mm} \text{(for all other $\ell$)},
\end{aligned}
\end{equation}
and
\begin{equation}
M_{\ell,ij} = \left(- \frac{40}{3} q_i \, \delta_{ij} + \frac{q_j}{q_i \bar{f}_i} \left[2 K^{\rm m}_{\ell,ij} - \frac{2}{9} q_i^2 q_j^2 \e^{-q_i} \left( 10 \delta_{\ell 0} -5 \delta_{\ell 1} +\delta_{\ell 2} \right) \right] \, w_j \right)
\label{scattering_matrix}
\end{equation}
is the scattering matrix encapsulating the collision kernels.
At very early times, the $G_{\ell,i}$-term is much smaller than the interaction term, and equation~\eqref{matrix_equation} becomes a homogeneous matrix equation with solution 
\begin{align}
\fett{\Psi}_{\ell}^h &= \sum_k c_k \, \fett{v}_{k} \exp{ \left( \lambda_k \frac{2\mathrm{N} T_{\nu,0}^5 G_{\text{eff}}^2}{(2\pi)^{3}} \int \frac{\mathrm{d} \eta}{a^4} \right)} \, ,\label{homogeneous_solution} \\
&\approx  \sum_k c_k \, \fett{v}_{k} \exp{ \left( \lambda_k \frac{2\mathrm{N} T_{\nu,0}^5 G_{\text{eff}}^2}{(2\pi)^{3} a_\text{ini}^4} \left(\eta - \eta_\text{ini} \right) \right)}
\label{homogeneous_solution_approx}
\end{align}
where we have expanded the integral around the initial time in the second line.
The coefficients $c_k$ are fixed by the initial conditions, and $\lambda_k$ and $\fett{v}_{k}$  denote respectively the $k$th eigenvalue and eigenvector of $M_{\ell,ij}$. 

\begin{figure}[t]
\includegraphics[width=\linewidth]{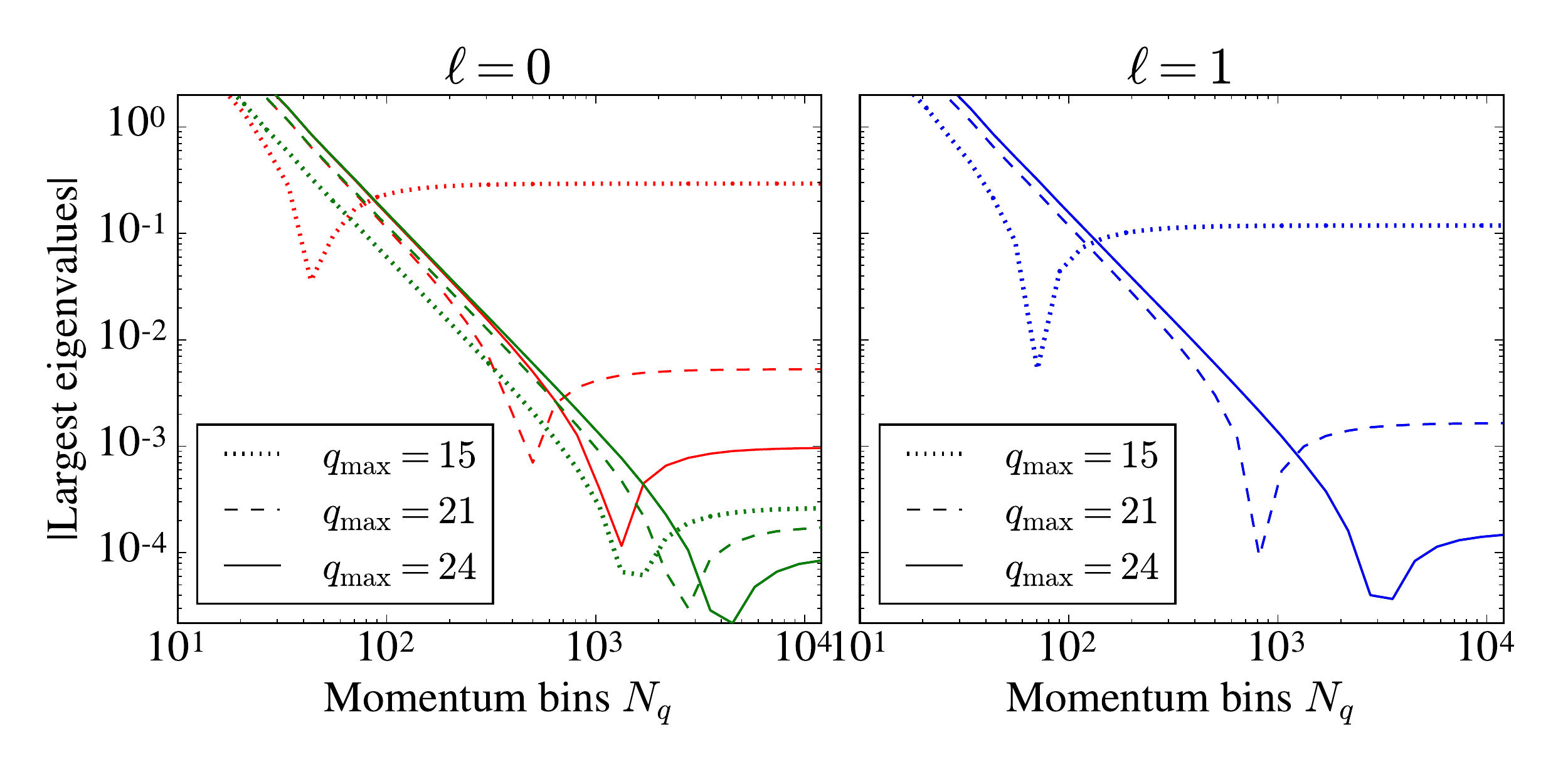}
\caption{\textit{Left:} Magnitudes of the two largest eigenvalues (red and green) at $\ell=0$ as functions of the number of momentum bins $N_{q}$ for $q_{\text{max}}=15$ (dotted), $q_{\text{max}}=21$ (dashed), and $q_{\text{max}}=24$ (solid). \textit{Right:} Same as the left panel, but at $\ell=1$, and only the largest eigenvalue (blue).}
\label{eigenvalues}
\end{figure}

An intriguing feature of the scattering matrix $M_{\ell,ij}$ in equation~\eqref{scattering_matrix} is that, for small numbers of momentum bins~$N_q$, some eigenvalues at $\ell=0$ and $\ell =1$ can become positive---up to two at $\ell=0$ and one at $\ell=1$.
From equation \eqref{homogeneous_solution_approx} it is clear that any positive eigenvalue~$\lambda_k$ will cause $\fett{\Psi}_{\ell}^h$ to grow exponentially with conformal time $\eta$ for some time after the initial time, leading to a numerical instability for large values of $G_{\text{eff}}$. For smaller values, the system remains stable since the exact solution of equation~\eqref{homogeneous_solution} remains bounded. Such a behaviour is not physical, since the effect of the scattering should always be to damp the perturbation. Indeed, the positive eigenvalues at $\ell=0,1$ for small $N_q$  values are an artefact of a finite grid size.  This is demonstrated in figure \ref{eigenvalues}, where we show the absolute values of the largest eigenvalues of the matrices $M_{0,ij}$ and $M_{1,ij}$ as functions of $N_q$.  As we increase $N_q$ the eigenvalues decrease in magnitude, turn negative (the turnaround points correspond to the troughs in the curves), and finally reach their asymptotic values. 

%%%%%%%%
%%%%%%%%
%%%%%%%%
%%%%%%%

Clearly, to remove all positive eigenvalues from the $\ell=0,1$ equations would require several thousands of momentum bins. Needless to say, such a large number of momentum bins is not numerically feasible in a Boltzmann code.   
Furthermore, the values to which the eigenvalues asymptote  depend on the chosen momentum cut-off scale $q_{\text{max}}$ in the discretisation,  illustrated also in figure~\ref{eigenvalues} by the sets of solid/dashed/dotted curves representing different values of $q_{\rm max}$.  This, however, may be used to our advantage: since the magnitudes of the asymptotic eigenvalues appear to decrease with increasing $q_{\text{max}}$, we may conclude that 
\begin{equation}
\underset{N_q, q_{\text{max}} \rightarrow \infty}{\text{lim}} \lambda_{\text{max}} =0 \,,
\end{equation}
and on this basis implement the following routine:
\begin{enumerate}
\item We calculate the eigenvalues of the scattering matrix $M_{\ell,ij}$ \eqref{scattering_matrix};
\item set the positive eigenvalues to their asymptotic ($N_q, q_{\text{max}} \rightarrow \infty$) values, i.e., zero; and 
\item derive a corrected scattering matrix, i.e., $M'_{\ell,ij}=P_{ik} D'_{kl} P^{-1}_{lj}$, where $D'_{kl}$ is the diagonal matrix with the corrected eigenvalues as diagonal entries, and $P_{ik}$ is the matrix that diagonalises $M_{\ell,ij}$.
\end{enumerate}

To make the system numerically stable it is furthermore very important to choose a sufficiently large value for $q_{\text{max}}$, as too small a value of $q_{\text{max}}$ would cause the results to be dependent on~$N_q$. We have tested the above routine against different choices of $N_q$ and  the cut-off momenta $q_{\text{min}}$ and $q_{\text{max}}$, and found that choosing $q_{\text{min}}=0.1$, $q_{\text{max}}=24$, and $N_{q}=24$ suffices to obtain numerically stable solutions of the exact Boltzmann hierarchy~\eqref{finalBoltzmann_mass}.
   
%%%%%%%%%%%%%%%%%%%%%%%%%%%%%%%%%%%%%%%%%%%%%%%%%%%%%%%%%

\subsection{Impact of interacting neutrinos on the CMB}
\label{Impact of interacting neutrinos on the CMB}

%%%%%%%%%%%
\begin{figure}[t] % there are no bottom figures in jcap
\centering
\includegraphics[width=\linewidth]{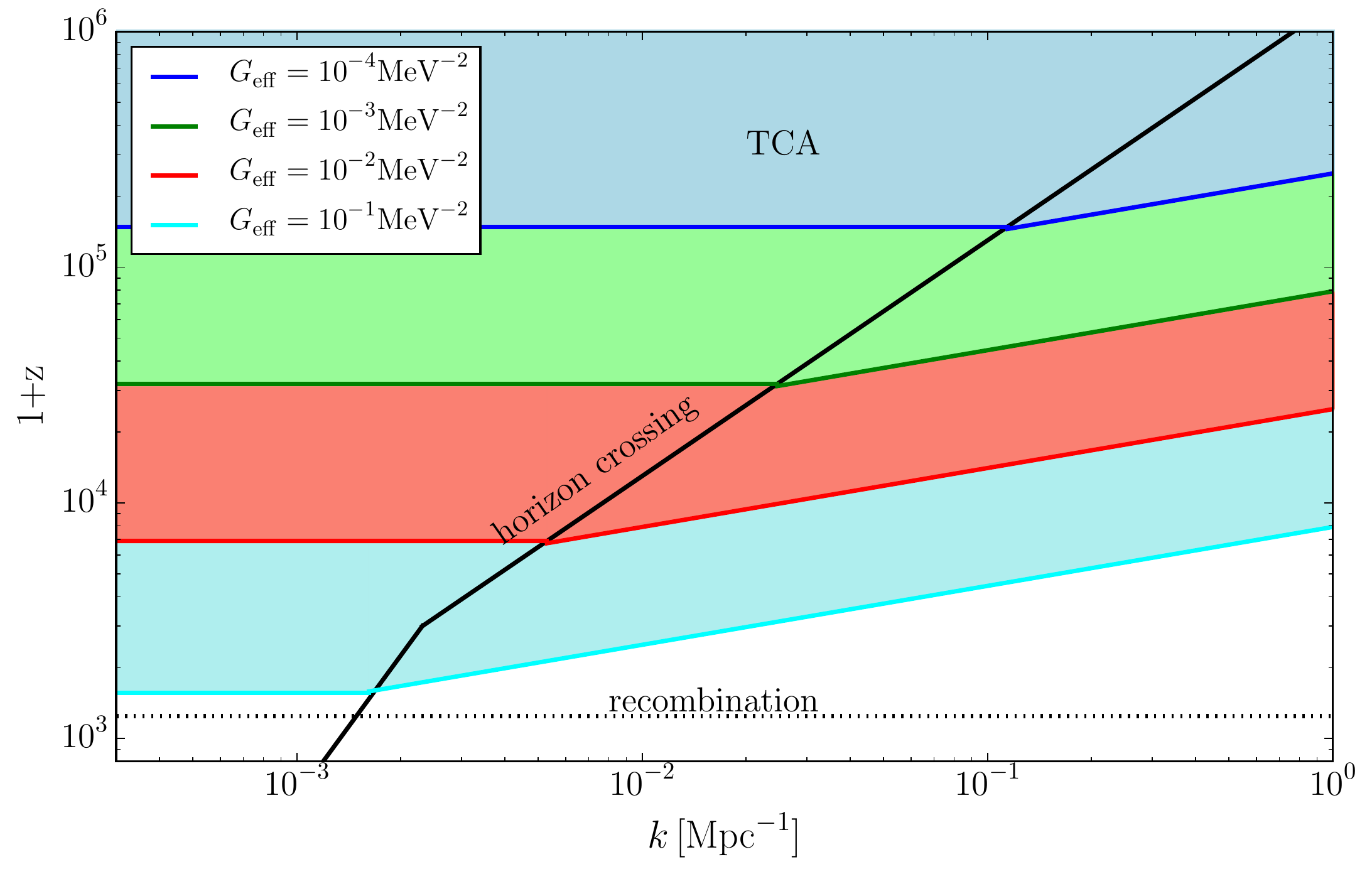}
\caption{Regions of validity of the tightly-coupled approximation (TCA) in the $(k,z)$-plane for various values of the effective coupling constant $G_{\text{eff}}$.  The black line represents $k = {\cal H}$, i.e., horizon crossing of a wavenumber~$k$, while the coloured lines correspond to either $|\dot{\tau}_\nu| = {\cal H}$ if a wavenumber is super-horizon $k < {\cal H}$, or $|\dot{\tau}_\nu| = k$ if a wavenumber is sub-horizon $k > {\cal H}$.  For a given $G_{\rm eff}$, the TCA is valid in the region above the corresponding coloured line.  In compiling this plot we have used $T_\nu = (4/11)^{1/3} T_\gamma$ for the neutrino temperature,  $H \simeq T_{\gamma}^2/m_{\text{Pl}}$ for the Hubble expansion rate at $z \gtrsim 3000$, and $H \simeq (\omega_{\rm cdm}+\omega_b)^{1/2} a^{-3/2}$, assuming the Planck $\Lambda$CDM best-fit values, at $z \lesssim 3000$.}
\label{TCAplot}
\end{figure}
%%%%%%%%%

Three phenomenological variables are relevant for the evolution of the neutrino perturbations, namely, the conformal Hubble rate 
${\cal H}=\dot a /a = a H$, the comoving wavenumber $k$, and the conformal interaction rate $|\dot{\tau}_\nu|= a G^2_{\text{eff}} T_{\nu}^5$. 
The tightly-coupled approximation (TCA) holds whenever the interaction rate dominates over the other two, i.e., $|\dot{\tau}_\nu| \gg {\cal H}, k$.
 In this case, multipole moments at $\ell \geq 2$ in the neutrino Boltzmann hierarchy are strongly suppressed, so that only the $\ell=0,1$~equations in~\eqref{finalBoltzmann_mass} or~\eqref{eq:cyr} need to  be kept.
Furthermore, as we show in appendix~\ref{Proof of number-, energy- and momentum conservation}, the collision terms of the exact Boltzmann 
hierarchy~\eqref{finalBoltzmann_mass} vanish at $\ell=0$ and $\ell=1$ when integrated over momentum.  This leaves 
 \begin{equation}
 \begin{aligned}
 \dot{\delta}_{\nu} &= -\frac{4}{3} \theta_\nu- \frac{2}{3}\dot{h} \, ,  \\ 
% \dot{\theta}_{\nu}&=k^2 \left(\frac{1}{4} \delta_\nu- \sigma_\nu \right) \, 
 \dot{\theta}_{\nu}&= \frac{1}{4} \, k^2 \delta_\nu \, 
 \end{aligned}
\label{TCA_equations} 
\end{equation}
as the only nontrivial equations of motion for the neutrino perturbations.   In other words, interacting neutrinos can be described as a perfect fluid within the TCA.
Figure~\ref{TCAplot} summarises the regions of validity of the TCA in the $(k,z)$-plane for a range of $G_{\rm eff}$~values.

%%%%%%%%%%%%
\begin{figure}[t]
\centering
\includegraphics[width=0.9\linewidth]{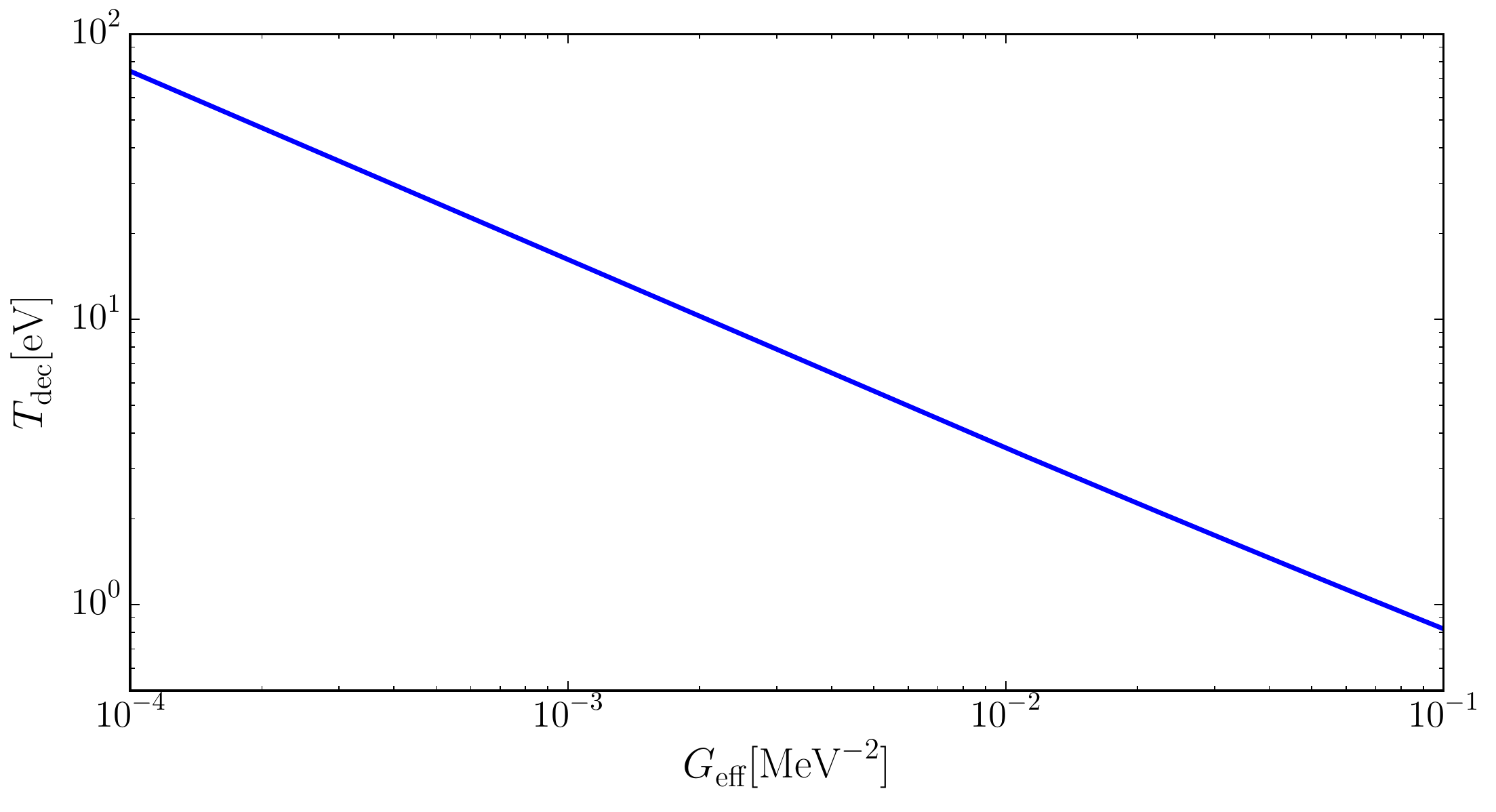}
\caption{Neutrino decoupling temperature, as defined in equation~(\ref{eq:Tdec}), as function of the effective coupling constant $G_{\text{eff}}$.  See figure~\ref{TCAplot} caption for the modelling of the Hubble expansion rate.}
\label{fig:decouplingtemp}
\end{figure}
%%%%%%%%%%%%%%%%%%%%%%%%%%%%

Holding $G_{\rm eff }$ fixed, we see in figure~\ref{TCAplot} that the TCA always fails first at large wavenumbers, followed by progressively smaller $k$ values until $k = {\cal H}$.  Beyond this point, on super-horizon scales, the TCA formally fails for all $k < {\cal H}$ at the same time.  It is therefore useful  to define a neutrino decoupling temperature, $T_{\rm dec}$,  as the temperature of the photons at which 
\begin{equation}
\label{eq:Tdec}
|\dot{\tau}_\nu (T_{\rm dec})| = {\cal H} (T_{\rm dec}).
\end{equation} 
Evaluated explicitly for the radiation-domination era, we find
\begin{equation}
T_{\text{dec}}\simeq 7.66 \times 10^{-2} \left( \frac{{\rm MeV}^{-2}}{G_{\text{eff}}} \right)^{2/3}\, \text{eV} = 0.2 \left( \frac{2.03 \times10^{10}\, G_{\rm F}}{G_{\text{eff}}} \right)^{2/3}\, \text{eV}, \quad ({\rm RD})
\label{T_dec}
\end{equation} 
where the numerical prefactors have been obtained using $H  \simeq T_{\gamma}^2/m_{\text{Pl}}$, and $T_\nu/T_\gamma = (4/11)^{1/3}$.  Of course, neutrinos with different momenta decouple at different times.  However, the measure~(\ref{eq:Tdec}) is still indicative of the average behaviour of the thermal ensemble (or, equivalently, the behaviours of momentum modes around $q \sim 1$).   
Figure~\ref{fig:decouplingtemp} shows $T_{\rm dec}$ as a function of the effective coupling constant $G_{\rm eff}$. Note that the transition from radiation to matter domination in principle produces a change in the slope around $T \sim 1$~eV; the change is however nearly impossible to resolve in the plot.

\begin{figure}[t]
\includegraphics[width=\linewidth]{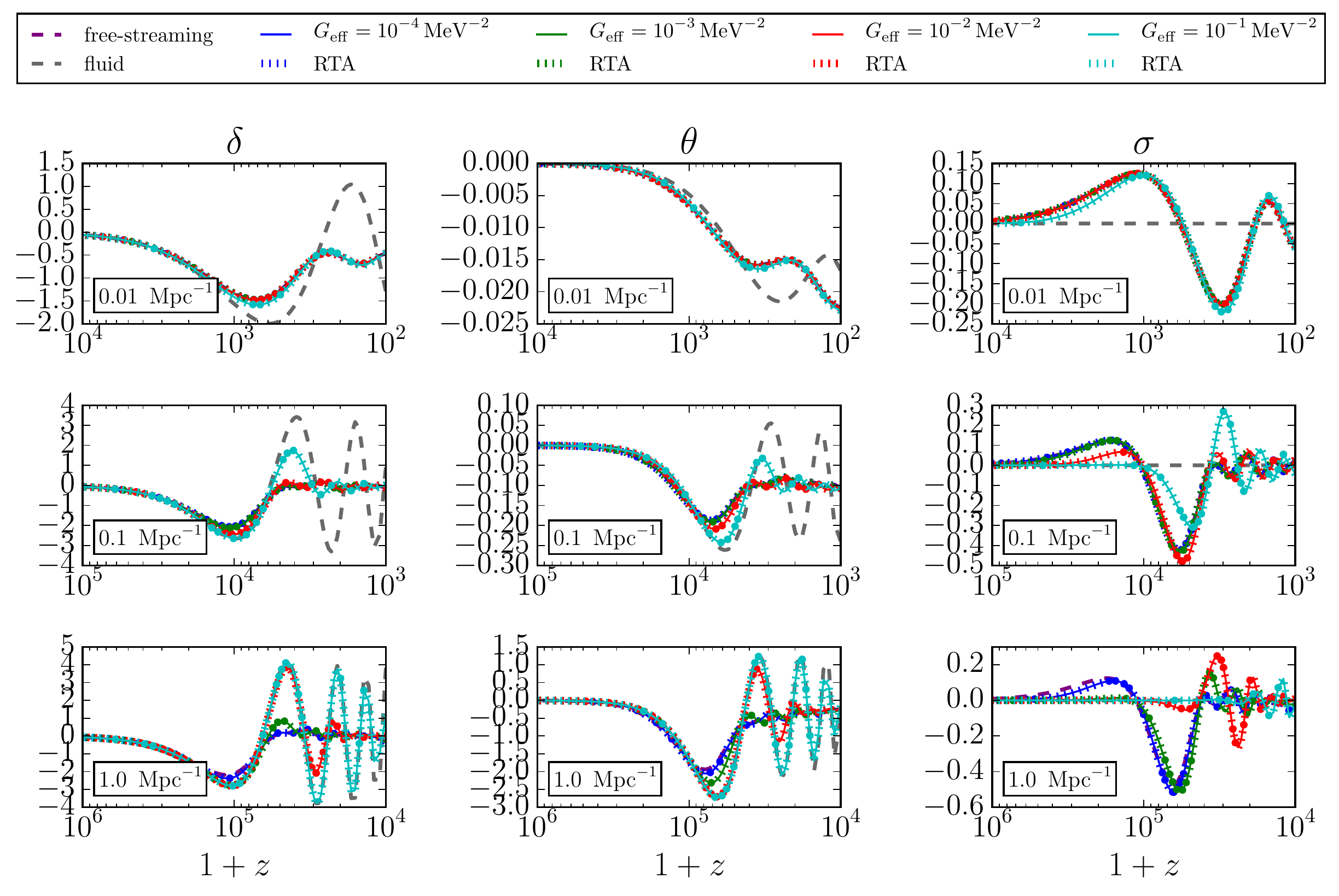}
\caption{Neutrino energy density contrast~$\delta$ (left), velocity divergence $\theta$~(middle), and shear stress~$\sigma$ (right) at $k=0.01$~Mpc$^{-1}$ (top), $0.1$~Mpc$^{-1}$ (middle), and $1.0$ Mpc$^{-1}$~(bottom)  for different values of the effective coupling constant $G_{\text{eff}}$ (in units of $\mathrm{MeV}^{-2}$), computed from both the exact Boltzmann hierarchy (solid) and the separable ansatz/RTA (short dashes).  The free-streaming and fluid limits are also shown for comparison.}
\label{perturbed}
\end{figure}

Figure~\ref{perturbed} shows, following the notation of~\cite{Ma:1995ey},  the neutrino energy density contrast~$\delta$, velocity divergence~$\theta$, and shear stress~$\sigma$ at various wavenumbers, computed using our implementation of the exact Boltzmann hierarchy~\eqref{finalBoltzmann_mass} in {\sc class}, for different values of the effective coupling constant~$G_{\text{eff}}$.  For ease of comparison, solutions in the standard free-streaming limit and the fluid limit represented by equation~(\ref{TCA_equations})  also appear in the figure.   In all cases, apart from the parameters that describe the neutrino interaction, all other cosmological model parameters have been set to their Planck $\Lambda$CDM best-fit values.

%%%%%%%%%%%%%
%%%%%%%%%%

 As expected, for  very large values of $G_{\rm eff}$, the perturbed quantities essentially track the fluid solutions at early times.  This fact is particularly well illustrated by the $k =1.0~{\rm Mpc}^{-1}$ case, where it is also apparent that  the larger the value of $G_{\rm eff}$, the longer the tracking period.  
When the conditions for TCA can no longer be satisfied, tracking ceases, accompanied by the generation of shear stress while power is being transferred to the higher multipoles of the Boltzmann hierarchy; the perturbed quantities eventually evolve to the standard free-streaming solutions.
Finally, we also test in figure~\ref{perturbed} the separable ansatz/RTA~\eqref{eq:cyr} against the exact Boltzmann hierarchy  (solid lines vs short dashes):  the agreement between the two approaches is, for all tested $G_{\rm eff}$ values and redshifts, very good.

%%%%%%%%%%%%%
\begin{figure}[t]
\centering
\includegraphics[width=\textwidth]{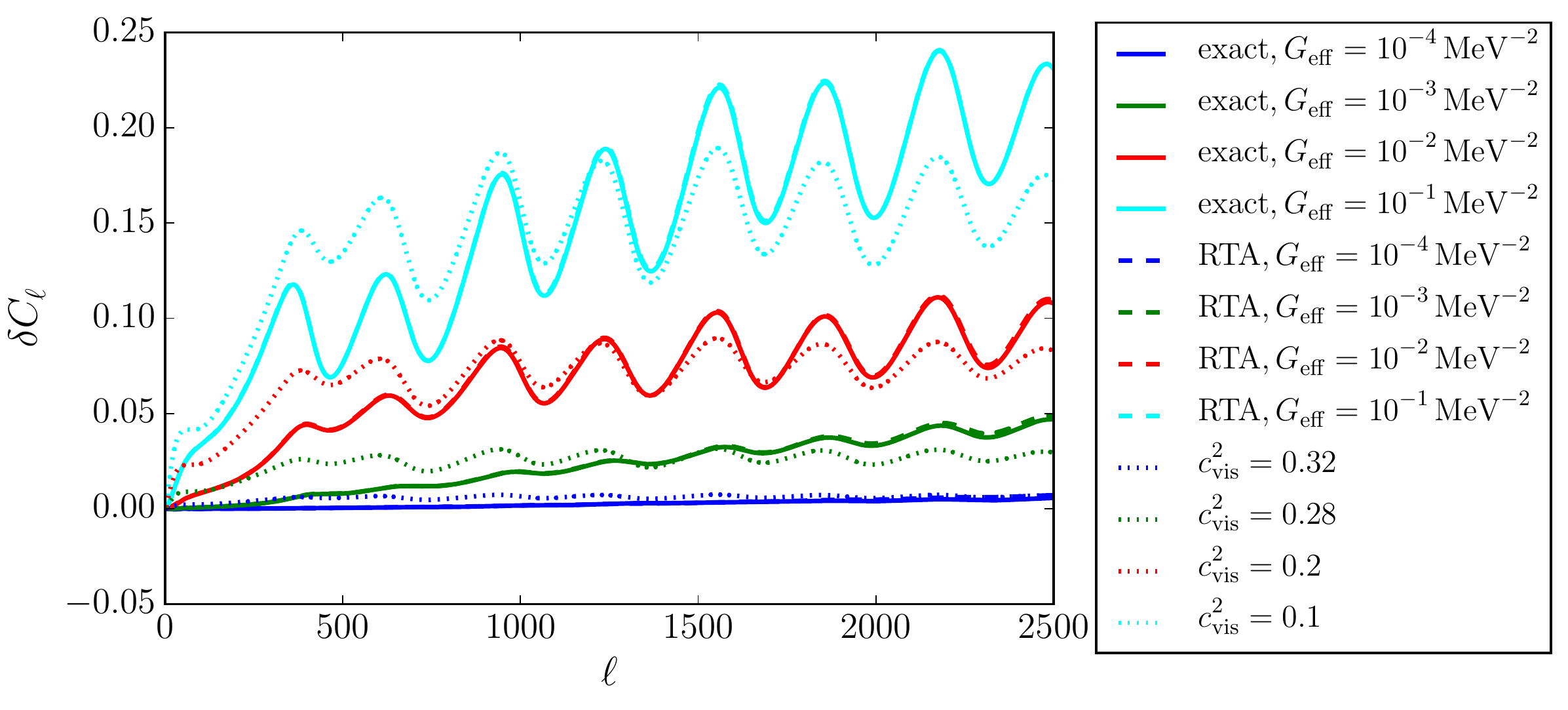}
\caption{Relative difference in the temperature anisotropy spectrum, $\delta C_{\ell}$, computed using the exact Boltzmann hierarchy~\eqref{finalBoltzmann_mass}, in comparison with solutions to the separable ansatz/RTA (dashed), and the $(c_{\text{eff}}^2,c_{\text{vis}}^2)$-parameterisation (dotted).}
\label{deltaCls}
\end{figure}
%%%%%%%%%%%

Turning now to the impact of interacting neutrinos on the CMB anisotropies, figure~\ref{deltaCls} shows  the relative difference in the temperature power spectrum, 
\begin{equation}
\delta C_{\ell} = \frac{C^{\text{int}}_{\ell}-C^{\Lambda \rm CDM}_{\ell}}{C^{\Lambda \rm CDM}_{\ell}},
\end{equation}
between an interacting scenario (``int'') and the standard free-streaming case (``$\Lambda$CDM'').  As expected, the larger the value of $G_{\rm eff}$, the greater the relative difference~$\delta C_\ell$ engendered by neutrino interactions.  Modulo the oscillations, for a given $G_{\rm eff}$ the greatest enhancement occurs on small angular scales (or large values of $\ell$).  For coupling constants as large as, e.g.,  $G_{\rm eff} = 10^{-1}~{\rm MeV}^{-2}$, neutrino interactions increase the CMB temperature power by almost 25\% at $\ell \sim 2000$.  Conversely,  the case of $G_{\rm eff} = 10^{-4}~{\rm MeV}^{-2}$ barely registers a 1\% difference at $\ell \sim 2500$.  This also sets the ballpark for the scale of $G_{\rm eff}$ that can be probed using CMB measurements.

Comparing the relative differences computed using (a) our exact Boltzmann hierarchy~\eqref{finalBoltzmann_mass}, (b) the separable ansatz/RTA~\eqref{eq:cyr},  and (c) the $(c_{\text{eff}}^2,c_{\text{vis}}^2)$-parameterisation~\eqref{ceff_cvis_parametrisation}, we see in figure~\ref{deltaCls} that our exact approach and the separable ansatz/RTA yield essentially the same result.%
\footnote{For internal consistency we have used the model-dependent coefficients $\alpha_\ell$ of equation~\eqref{alphas} together with the separable ansatz/RTA~(\ref{eq:cyr}).  These coefficients are $\ell$-dependent, and range from 0.40 to 0.48.  However, we could equally have set them all to 1.0: save for a $\sim 40$\% difference in the definition of $G_{\rm eff}$, these two choices of $\alpha_\ell$s have an almost identical impact on the CMB anisotropies.}
This comes as no surprise given the remarkable agreement between our exact approach and the separable ansatz already demonstrated in figure~\ref{perturbed}.
We conclude therefore that the separable ansatz/RTA  suffices to model the CMB phenomenology of neutrinos interacting via an effectively four-fermion vertex.

In contrast, modulo the oscillations, the $(c_{\text{eff}}^2,c_{\text{vis}}^2)$-parameterisation produces a fairly uniformly enhanced temperature power spectrum at $\ell \gtrsim 200$; no matter what $c_{\rm vis}$ value we take as an input, the model clearly does not reproduce the full scale dependence of the exact approach or of the separable ansatz/RTA.
We therefore conclude that the $(c_{\text{eff}}^2,c_{\text{vis}}^2)$-parameterisation has neither a formal nor a phenomenological interpretation in the context of particle scattering, and hence should be avoided.

%%%%%%%%%%%%%%%%%%%%%%%%%%%%%%%%%%%%%%%%%%%%%%%%%%%%%%%%%

\section{Constraints on the effective coupling constant~\texorpdfstring{$G_{\text{eff}}$}{Geff}}
\label{Constraints on the neutrino coupling}

Given the demonstrated accuracy of the RTA in the massive scalar limit, we shall be working within this approximation in the following when testing neutrino self-interactions against cosmological observations in a Markov Chain Monte Carlo (MCMC) analysis.

\subsection{Data sets}

 We use the following data sets in our analysis:
\begin{description}
\item[TT] Temperature power spectrum and low-$\ell$ polarisation from the Planck 2015 data~\cite{Ade:2015xua}.
\item[CMB] Same as TT, but including also the $E$-polarisation power spectrum (EE) and $E$-polarisation cross-power spectrum (TE) from Planck 2015~\cite{Ade:2015xua}.
\item[BAO] Baryon acoustic oscillation peak scale as measured by 6DF~\cite{Beutler:2011hx}, BOSS LOWZ and CMASS~\cite{Anderson:2013zyy,Cuesta:2015mqa}, and the SDSS Main Galaxy Sample~\cite{Ross:2014qpa}.
\item[HST] Measurements of the local Hubble expansion rate by~\cite{Riess:2016jrr}.
\end{description}
From these data sets we construct four different data combinations, all containing the TT and the BAO measurements: TT+BAO, CMB+BAO, TT+BAO+HST, CMB+BAO+HST.

%%%%%%%%%%%

\subsection{Method}
Using the MCMC engine Monte Python~\cite{Audren:2012wb}, we perform an MCMC analysis of two cosmological models containing self-interacting neutrinos:
\begin{enumerate}
\item SI$\nu$: Comprises three species of massless neutrinos all interacting with strength $G_\text{eff}$. The parameters of the model are 
\begin{equation}
\label{eq:parameters}
\left\{ \omega_\text{cdm}, \, \omega_b, \, 100\theta_s, \, \ln \left(10^{10}A_s \right), \,  n_s, \, z_\text{reio}, \, \log_{10}\left(G_\text{eff}/{\rm MeV}^{-2} \right) \right\},
\end{equation}
where $\theta_s$ is the sound horizon at recombination, and the other symbols carry their usual meanings.

\item  SI$\nu$+$N_\text{eff}$: Same as SI$\nu$, but the number of self-interacting neutrino species, $N_{\rm eff}$, is allowed to vary, thereby expanding the model parameter list to
\begin{equation}
\label{eq:parameters2}
\left\{ \omega_\text{cdm}, \, \omega_b, \, 100\theta_s, \, \ln \left(10^{10}A_s \right), \,  n_s, \, z_\text{reio}, \, \log_{10}\left(G_\text{eff}/{\rm MeV}^{-2} \right), N_{\rm eff} \right\}.
\end{equation}
\end{enumerate}
We assume flat priors on all of these parameters, and restrict the coupling constant to the range $\log_{10}(G_\text{eff}/{\rm MeV}^{-2}) \in  [-5,0]$.  Convergence of the Markov Chains is determined by the Gelman--Rubin convergence criterion $R< 0.01$.

For large couplings $G_{\rm eff}$ the system of equations becomes so stiff that even the implicit ODE-solver in \CLASS{} fails to evolve the system. This happens at $\log_{10}(G_\text{eff}/{\rm MeV}^{-2}) \gtrsim -0.1$. But even at $\log_{10}(G_\text{eff}/{\rm MeV}^{-2}) \sim -1.0$ \CLASS{} already becomes prohibitively slow.  Therefore,  in order to generate MCMC samples in a finite amount of time, we restrict the maximum number of steps allowed in the ODE-solver to $10^5$. This restriction does not appear to have affected the SI$\nu$-chains significantly, but the $\text{SI}\nu+N_\text{eff}$ chains have been cut off at large $N_\text{eff}$ values and  $\log_{10}(G_\text{eff}/{\rm MeV}^{-2}) \gtrsim -1.0$.

\begin{figure}[t]
\includegraphics[width=\textwidth]{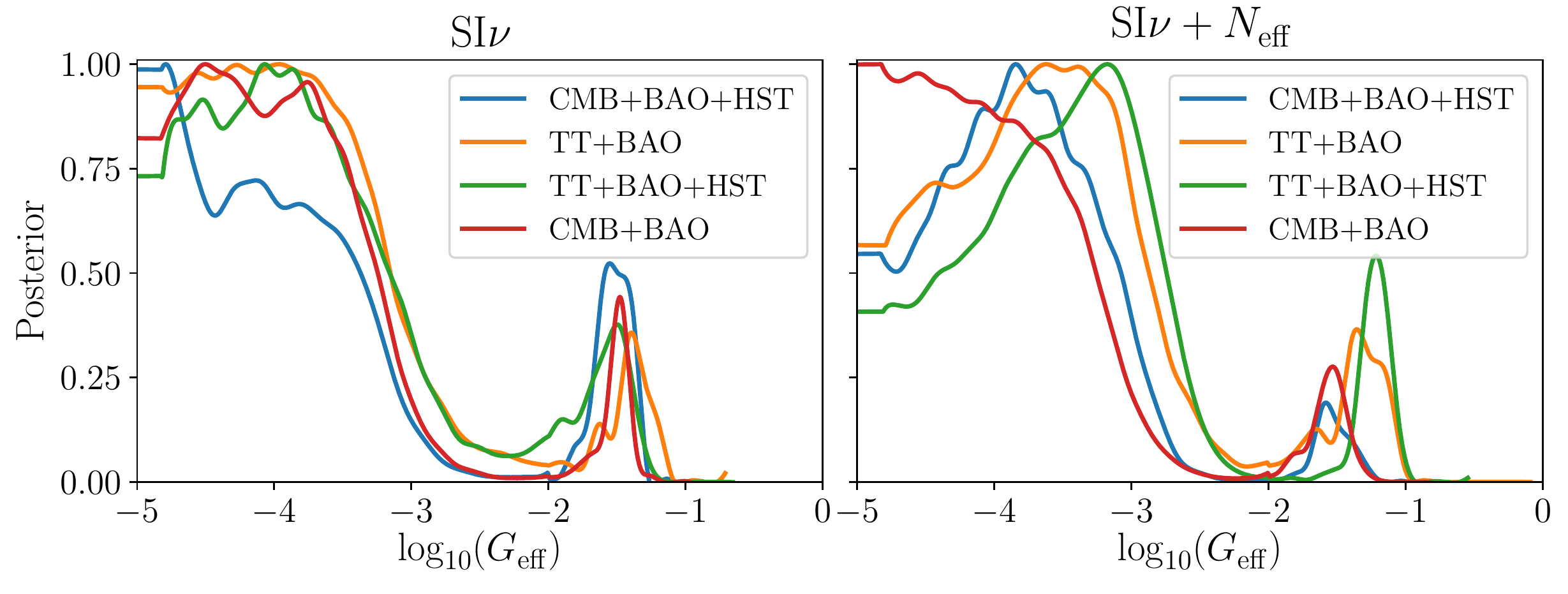}
\caption{1d marginalised posterior distribution of the effective coupling constant $G_{\text{eff}}$ (in units of  $\text{MeV}^{-2}$), derived from various data  combinations, for a model with $3.046$ interacting neutrino species (left), and a model in which the number of interacting species, $N_{\text{eff}}$, is  a free parameter (right).}
\label{fig:GeffPosteriorsTot}
\end{figure}

\begin{figure}[t]
\includegraphics[width=\textwidth]{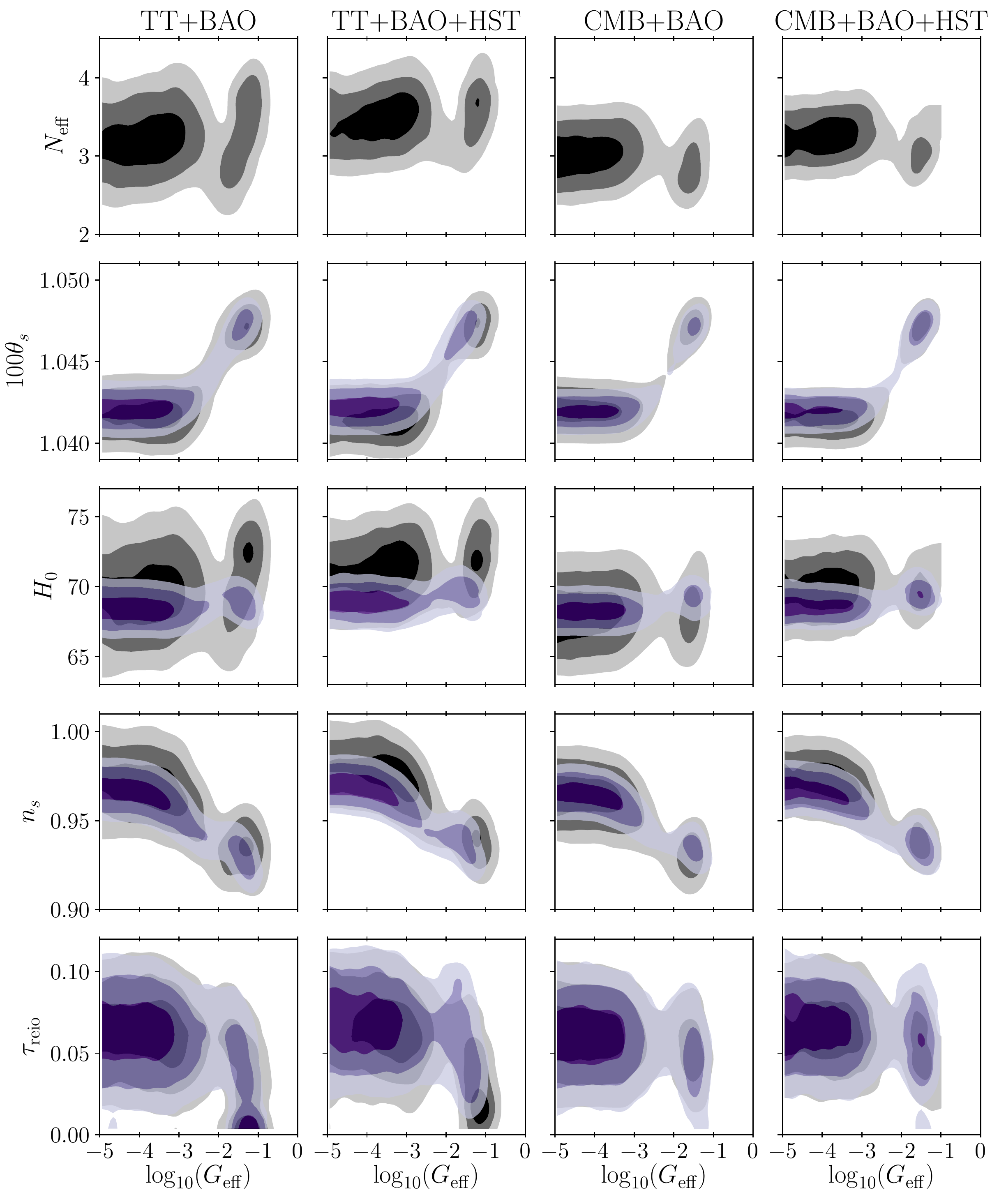}
\caption{2d marginalised posterior distributions of the parameters most affected by neutrino self-interactions, where the contours demarcate the 68\% and 95\% confidence regions,
 for SI$\nu$ (grey) and SI$\nu$+$N_{\text{eff}}$ (violet). The effective coupling constant $G_{\rm eff}$ is quoted here in units of ${\rm MeV}^{-2}$, and the Hubble parameter in units of km/s/Mpc.}
\label{fig:GeffCorrelations1}
\end{figure}

Looking at figures~\ref{fig:GeffPosteriorsTot} and~\ref{fig:GeffCorrelations1} , it is also clear that a small second peak exists, in addition to the main peak at $G_{\rm eff} \to 0$.
This second peak was first pointed out in~\cite{Cyr-Racine:2013jua} and tentatively in~\cite{Archidiacono:2013dua} in their analyses of the Planck temperature measurements; the statistical significance was however low.   Adding polarisation data and measurements of the local Hubble expansion rate substantially  drives up the statistical significance of the second peak.

%%%%%%%%
%%%%%%%%

%%%%%%%%%%
%%%%%%%%%

In order to properly sample the second peak, we increase the temperature of the chains by a factor  of three, i.e., we sample the rescaled distribution $P^{1/3}$. A further complication arises from the narrowness of  the peak, which necessitates a small bin width and forces us to turn off spline-smoothing of the posteriors. Instead, we use a simple moving average smoothing in the $\log_{10}(G_\text{eff}/{\rm MeV}^{-2}) < -2.0$ part of the posterior, effectively increasing the bin-size in this part of the distribution by a factor  of $3.5$. Similar problems arise in the construction of the 2D contours, since the two islands are connected by an isthmus: applying a Gaussian smoothing with too large a smoothing scale would change the plot qualitatively by disconnecting the islands, so we must use a small smoothing scale.

As is evident in figure~\ref{fig:GeffCorrelations1}, the two peaks are most well separated in the $\theta_s$ parameter.  We therefore divide up the parameter space in two modes, {\it M1} and {\it M2}, along a plane of constant~$\theta_s$ defined in table~\ref{tab:M1M2}, and consider the two modes independently.
%We could of course also have  defined the modes directly in the effective coupling constant $\log_{10}(G_\text{eff})$, which would have been more straightforward. However, our present approach allows us to get overlapping posteriors in the effective coupling constant 

%%%%%%%%%%%%%%%%
\begin{table}[t]
\centering
\begin{tabular}{l|ll}
& {\it M1}  & {\it M2} \\
\hline
$\text{SI}\nu$                       & $100 \theta_s \leq 1.0438$     & $100 \theta_s \geq 1.0438$ \\
$\text{SI}\nu+N_\text{eff}$  & $100 \theta_s \leq 1.046485$ & $100 \theta_s \geq 1.046485$
\end{tabular}
\caption{The {\it M1} and {\it M2} modes, defined in terms of the parameter ranges of the sound horizon $100 \theta_{s}$. }
\label{tab:M1M2}
\end{table}
%%%%%%%%%%%%%%%%%%%%%%%%%

%%%%%%%%%%

\subsection{Discussions}

\begin{figure}[t]
\includegraphics[width=14.8cm]{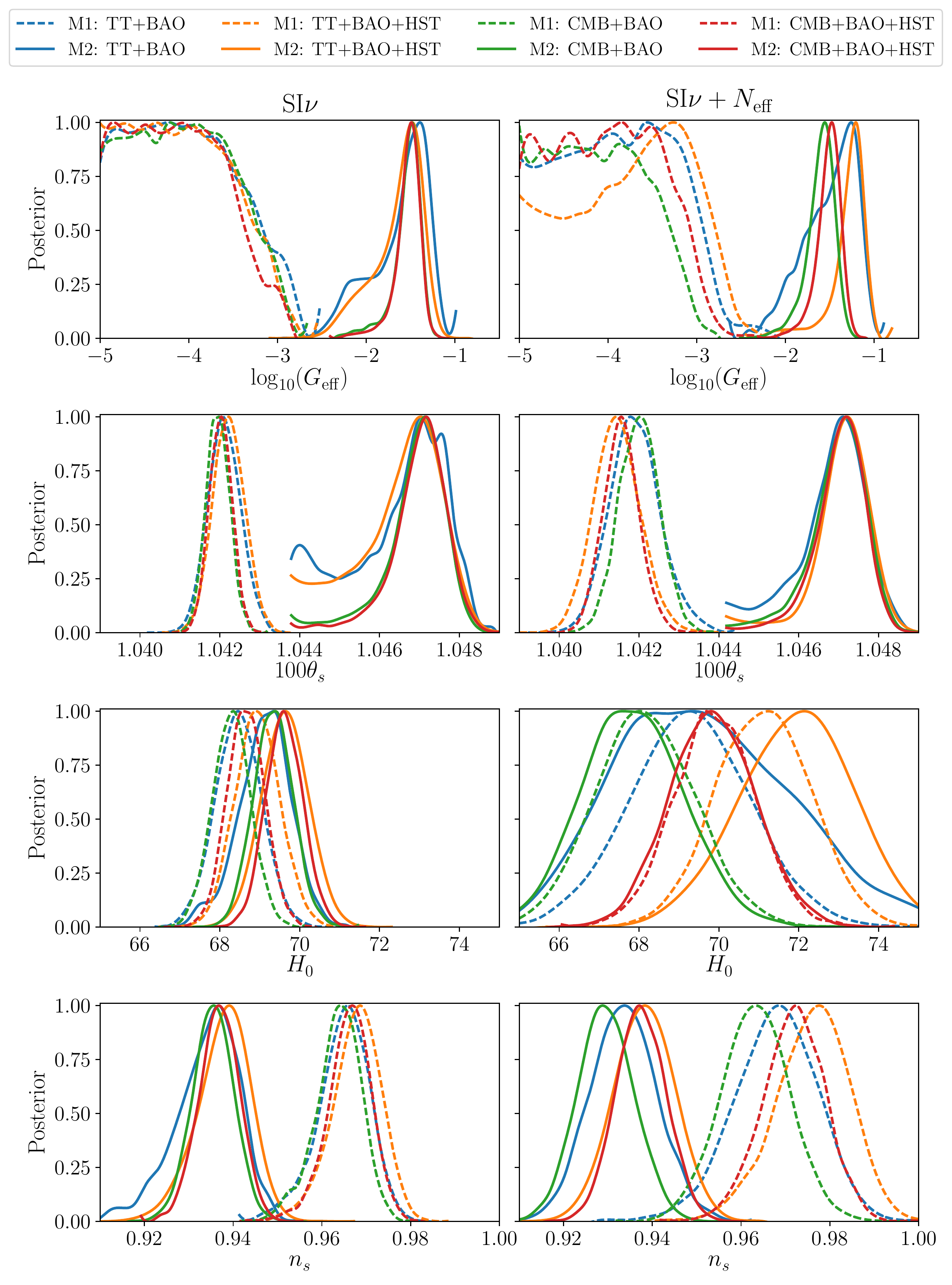}
\caption{1d posterior distributions of the parameters most affected by neutrino self-interactions.  The effective coupling constant $G_{\rm eff}$ is quoted in units of ${\rm MeV}^{-2}$, and the Hubble parameter in units of km/s/Mpc.}
\label{fig:AllPosteriors}
\end{figure}

Figure~\ref{fig:AllPosteriors} shows the 1d posteriors for selected parameters in each mode, for both the $\text{SI}\nu$ and the $\text{SI}\nu+N_\text{eff}$ model,  derived from four data combinations TT+BAO, TT+CMB+BAO, CMB+BAO, and CMB+BAO+HST.  In both models, the posterior of~$\log_{\text{10}}(G_{\text{eff}})$ clearly shows that {\it M2} is associated with large values of the effective coupling constant $G_{\rm eff}$, while {\it M1} corresponds to the non-interacting, free-streaming limit. Specifically, we find the constraints (see also tables~\ref{tab:parameters} and~\ref{tab:Neffparameters}):
\begin{equation}
\begin{aligned}
\log(G_{\rm eff}/{\rm MeV}^{-2}) & \lesssim -2.9~(95\%), \qquad \quad \quad \;\; (M1,{\rm CMB+BAO+HST}) \\
\log(G_{\rm eff}/{\rm MeV}^{-2}) & = -1.5 \pm 0.3~(68\%), \qquad (M2,{\rm CMB+BAO+HST}) \\
\end{aligned}
\end{equation}
to be representative of $\text{SI}\nu$, while for $\text{SI}\nu+N_\text{eff}$ the same data combination yields a very similar
\begin{equation}
\begin{aligned}
\log(G_{\rm eff}/{\rm MeV}^{-2}) & \lesssim -2.6~(95\%), \qquad \quad \quad \;\; (M1,{\rm CMB+BAO+HST}) \\
\log(G_{\rm eff}/{\rm MeV}^{-2}) & = -1.5 \pm 0.3~(68\%). \qquad (M2,{\rm CMB+BAO+HST}) \\
\end{aligned}
\end{equation}
In terms of the neutrino decoupling temperature (which can be read off figure~\ref{fig:decouplingtemp} given a $G_{\rm eff}$ value), the {\it M1} bound on $G_{\rm eff}$ translates approximately to a lowest decoupling temperature of $T \sim 10$~eV, while 
the {\it M2} peak corresponds to decoupling at $T \sim 2$~eV. 

\begin{figure}[t]
\centering
\includegraphics[width=0.9\linewidth]{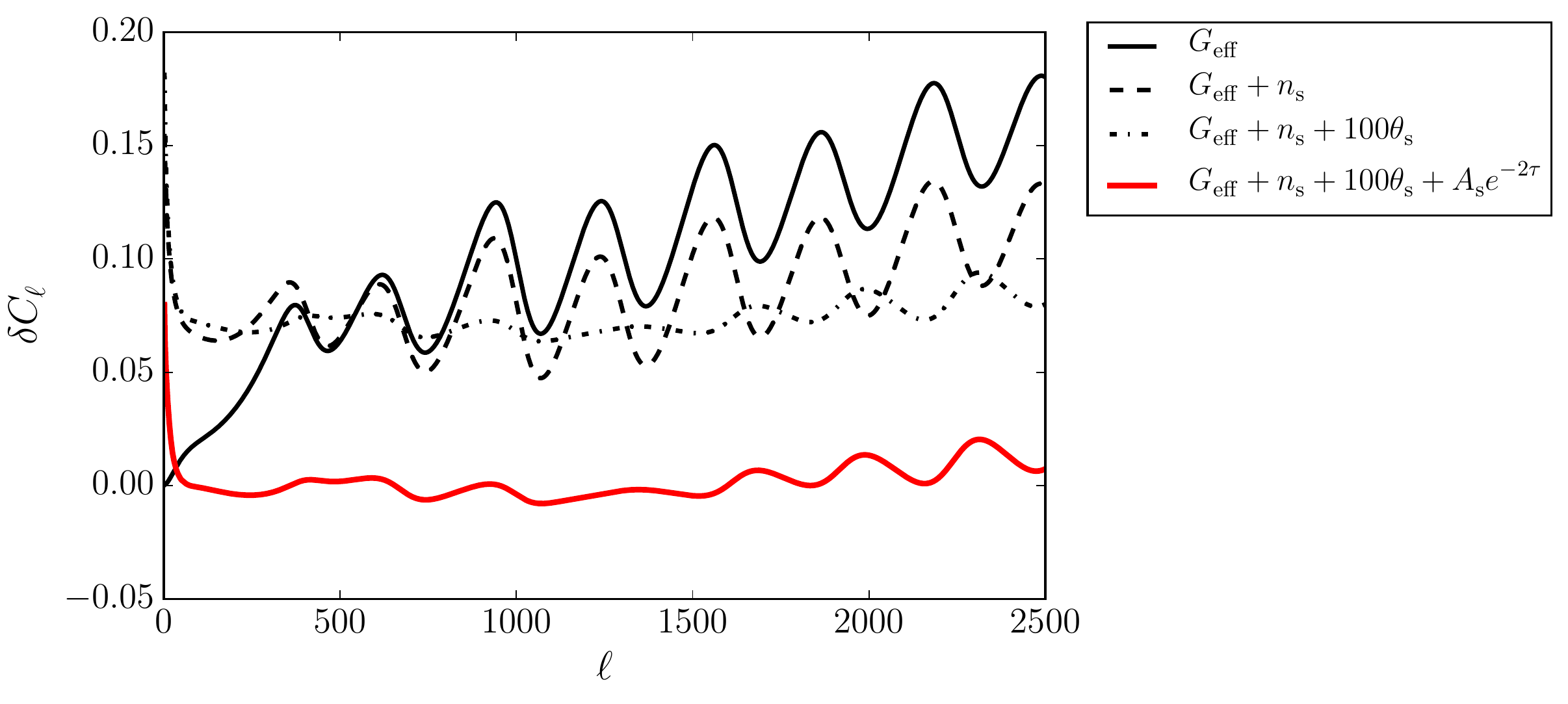}
\caption{Differences induced in the temperature anisotropy spectrum relative to the \textit{M1} best-fit  obtained by varying $G_{\rm eff}$ (solid black), $G_{\rm eff}+n_s$ (dashed), $G_{\rm eff}+n_s+100\theta_s$ (dot dash), and $G_{\rm eff}+n_s+100\theta_s+A_s e^{-2\tau}$ (solid red) to their respective best-fit values in the \textit{M2} mode.}
\label{parameter_degeneracy}
\end{figure}

Furthermore, as is evident in figure \ref{fig:AllPosteriors}, \textit{M2} is  accompanied by a lower spectral index~$n_s$ and a higher sound horizon $\theta_s$ compared with \textit{M1}.  This result can be understood from figure~\ref{parameter_degeneracy}, where we show the difference induced in the temperature anisotropy spectrum from variations of $G_{\rm eff}$, $n_s$, $\theta_s$, and $A_s e^{-2 \tau}$ relative to the \textit{M1} mode.  Firstly, 
the preference for a lower $n_s$ comes about because interacting neutrinos add power on small scales in a gradual manner that can be partially compensated by a redder spectral index. Secondly, the shift in the acoustic peaks due to the absence of neutrino anisotropic stress caused by a large~$G_{\mathrm{eff}}$ 
can be partially cancelled by  increasing $\theta_s$. The simultaneous variation of $G_{\rm eff}$, $n_s$, and $\theta_s$ therefore results in a temperature anisotropy spectrum that closely mimics the free-streaming~\textit{M1} mode, where the remaining offset can be easily touched up by adjusting the primordial fluctuation amplitude~$A_s$ and optical depth to reionisation $\tau$.  Note also that these parameter degeneracies are only approximate; this explains the bimodality of the posterior distribution, because some values of $G_{\mathrm{eff}}$ can be better compensated than others. See also the discussion in section~5.1 of~\cite{Lancaster:2017ksf}.

%%%%%%
%%%%%%%%

Observe that, in both models and especially $\text{SI}\nu$, the $\theta_s$-posterior in {\it M2} does not drop to zero at the separation boundary for the TT+BAO data combination, indicating that the significance of {\it M2} for this data combination is quite weak.   Adding polarisation and HST data drives the posterior to zero at the boundary, producing a much more distinct  {\it M2} peak.  Formally, the significance of  {\it M2} relative to {\it M1} may be quantified via the $\chi^2$-values at their respective best-fit points.   These are tabulated in the last rows of tables~\ref{tab:parameters} and~\ref{tab:Neffparameters}, for  $\text{SI}\nu$ and $\text{SI}\nu+N_\text{eff}$  respectively.  In both models and for all data combinations, we find  {\it M2} to be a worse fit to the data than  {\it M1}, albeit  only by a marginal $\Delta \chi^2 = 0.9$ for CMB+BAO+HST in the case of $\text{SI}\nu+N_\text{eff}$.

%%%%%%%%%%%%%%
\begin{table}
{\renewcommand{\arraystretch}{1.2}
\begin{tabular}{l|c|R{2.6cm}|R{2.6cm}|R{2.6cm}|R{2.6cm}|}
\multicolumn{2}{c}{} & \multicolumn{1}{c}{\makebox[2.6cm][c]{TT+BAO}} & \multicolumn{1}{c}{\makebox[2.6cm][c]{TT+BAO+HST}} & \multicolumn{1}{c}{\makebox[2.6cm][c]{CMB+BAO}} & \multicolumn{1}{c}{\makebox[2.6cm][c]{CMB+BAO+HST}}\\ \hline\hline 
\multirow{ 2}{*}{$\log_{10}(G_\mathrm{eff})$}& {\it M1} & $<-2.7~(95\%)$ & $<-2.8~(95\%)$ & $<-2.8~(95\%)$ & $<-2.9~(95\%)$\\
& {\it M2} & $-1.6_{-0.7}^{+0.5}$ & $-1.6_{-0.6}^{+0.5}$ & $-1.5_{-0.4}^{+0.3}$ & $-1.5_{-0.3}^{+0.3}$\\
\hline
\multirow{ 2}{*}{$100\theta_s$}& {\it M1} & $1.0421_{-0.0008}^{+0.0009}$ & $1.0422_{-0.0008}^{+0.0008}$ & $1.0420_{-0.0006}^{+0.0006}$ & $1.0420_{-0.0006}^{+0.0006}$\\
& {\it M2} & $1.0465_{-0.0025}^{+0.0017}$ & $1.0465_{-0.0025}^{+0.0016}$ & $1.0469_{-0.0019}^{+0.0014}$ & $1.0470_{-0.0016}^{+0.0014}$\\
\hline
\multirow{ 2}{*}{$H_0$}& {\it M1} & $68.5_{-1.2}^{+1.2}$ & $68.9_{-1.1}^{+1.1}$ & $68.3_{-1.0}^{+1.0}$ & $68.7_{-1.0}^{+1.0}$\\
& {\it M2} & $69.1_{-1.5}^{+1.3}$ & $69.6_{-1.2}^{+1.2}$ & $69.3_{-1.0}^{+1.0}$ & $69.6_{-0.9}^{+1.0}$\\
\hline
\multirow{ 2}{*}{$n_s$}& {\it M1} & $0.965_{-0.013}^{+0.012}$ & $0.968_{-0.012}^{+0.011}$ & $0.964_{-0.012}^{+0.010}$ & $0.966_{-0.011}^{+0.010}$\\
& {\it M2} & $0.934_{-0.016}^{+0.014}$ & $0.938_{-0.013}^{+0.012}$ & $0.935_{-0.010}^{+0.009}$ & $0.937_{-0.009}^{+0.009}$\\
\hline
\multirow{ 2}{*}{$\tau_\mathrm{reio}$}& {\it M1} & $0.06_{-0.03}^{+0.03}$ & $0.07_{-0.03}^{+0.03}$ & $0.06_{-0.02}^{+0.02}$ & $0.07_{-0.02}^{+0.02}$\\
& {\it M2} & $0.05_{-0.04}^{+0.03}$ & $0.06_{-0.03}^{+0.03}$ & $0.06_{-0.03}^{+0.03}$ & $0.06_{-0.02}^{+0.02}$\\
\hline
\multicolumn{2}{c|}{$\chi^2_{M2} - \chi^2_{M1}$} & 4.5 & 2.9 & 3.4 & 0.9
\end{tabular}
}
\caption{\label{tab:parameters}Mean values and 68\% credible regions of the cosmological parameters inferred from various data combinations, for the standard mode ({\it M1}) and the interacting mode ({\it M2}) in the $\text{SI}\nu$ model. The effective coupling constant $G_{\rm eff}$ is quoted in units of ${\rm MeV}^{-2}$, and the Hubble parameter in units of km/s/Mpc.}
\end{table}
%%%%%%%%%%

%%%%%%%%%%
\begin{table}
{\renewcommand{\arraystretch}{1.2}
\begin{tabular}{l|c|R{2.6cm}|R{2.6cm}|R{2.6cm}|R{2.6cm}|}
\multicolumn{2}{c}{} & \multicolumn{1}{c}{\makebox[2.6cm][c]{TT+BAO}} & \multicolumn{1}{c}{\makebox[2.6cm][c]{TT+BAO+HST}} & \multicolumn{1}{c}{\makebox[2.6cm][c]{CMB+BAO}} & \multicolumn{1}{c}{\makebox[2.6cm][c]{CMB+BAO+HST}}\\ \hline\hline 
\multirow{ 2}{*}{$\log_{10}(G_\mathrm{eff})$}& {\it M1} & $<-2.3~(95\%)$ & $<-2.4~(95\%)$ & $<-2.9~(95\%)$ & $<-2.6~(95\%)$\\
& {\it M2} & $-1.5_{-0.6}^{+0.5}$ & $-1.3_{-0.6}^{+0.3}$ & $-1.6_{-0.4}^{+0.3}$ & $-1.5_{-0.3}^{+0.3}$\\
\hline
\multirow{ 2}{*}{$N_\mathrm{eff}$}& {\it M1} & $3.2_{-0.5}^{+0.5}$ & $3.4_{-0.4}^{+0.4}$ & $3.0_{-0.4}^{+0.4}$ & $3.3_{-0.3}^{+0.3}$\\
& {\it M2} & $3.1_{-0.7}^{+0.8}$ & $3.5_{-0.6}^{+0.5}$ & $2.8_{-0.4}^{+0.4}$ & $3.1_{-0.3}^{+0.3}$\\
\hline
\multirow{ 2}{*}{$100\theta_s$}& {\it M1} & $1.0419_{-0.0013}^{+0.0014}$ & $1.0415_{-0.0012}^{+0.0012}$ & $1.0420_{-0.0010}^{+0.0010}$ & $1.0416_{-0.0009}^{+0.0009}$\\
& {\it M2} & $1.0468_{-0.0021}^{+0.0016}$ & $1.0471_{-0.0018}^{+0.0015}$ & $1.0470_{-0.0016}^{+0.0014}$ & $1.0470_{-0.0014}^{+0.0013}$\\
\hline
\multirow{ 2}{*}{$H_0$}& {\it M1} & $69.3_{-3.1}^{+3.1}$ & $71.1_{-2.4}^{+2.4}$ & $68.2_{-2.5}^{+2.6}$ & $69.9_{-2.1}^{+2.1}$\\
& {\it M2} & $69.6_{-4.2}^{+4.5}$ & $71.9_{-3.1}^{+2.9}$ & $67.9_{-2.6}^{+2.6}$ & $69.8_{-2.2}^{+2.2}$\\
\hline
\multirow{ 2}{*}{$n_s$}& {\it M1} & $0.968_{-0.020}^{+0.019}$ & $0.976_{-0.018}^{+0.017}$ & $0.963_{-0.016}^{+0.016}$ & $0.972_{-0.015}^{+0.015}$\\
& {\it M2} & $0.934_{-0.015}^{+0.015}$ & $0.938_{-0.014}^{+0.014}$ & $0.930_{-0.013}^{+0.013}$ & $0.937_{-0.012}^{+0.012}$\\
\hline
\multirow{ 2}{*}{$\tau_\mathrm{reio}$}& {\it M1} & $0.06_{-0.03}^{+0.03}$ & $0.07_{-0.03}^{+0.03}$ & $0.06_{-0.02}^{+0.02}$ & $0.07_{-0.02}^{+0.02}$\\
& {\it M2} & $0.05_{-0.03}^{+0.03}$ & $0.04_{-0.04}^{+0.03}$ & $0.06_{-0.03}^{+0.02}$ & $0.06_{-0.03}^{+0.03}$\\
\hline
\multicolumn{2}{c|}{$\chi^2_{M2} - \chi^2_{M1}$} & 4.9 & 3.5 & 2.6 & 1.8
\end{tabular}
}
\caption{\label{tab:Neffparameters} Same as table~\ref{tab:parameters}, but for the extended $\text{SI}\nu+N_\text{eff}$ model.}
\end{table}
%%%%%%%%%%

It is interesting to note in table~\ref{tab:parameters} that, before the addition of HST data, the inferred value of the Hubble parameter in  {\it M2} of  $\text{SI}\nu$  is approximately $1~\text{km}/{\rm s}/\text{Mpc}$ larger than in  {\it M1} of the same model. On its own this result is not enough to resolve the tension in $H_0$ between CMB and local measurements;  it does however invite one to speculate if allowing  $N_\text{eff}$ to vary at the same time would further increase $H_0$ to the HST value. However, as is evident in figure~\ref{fig:AllPosteriors} (and also in table~\ref{tab:Neffparameters}), allowing $N_\text{eff}$ to vary as well merely serves to draw  the $H_0$-posteriors of {\it M1} and {\it M2} closer to one another; there is no accompanying increase in the inferred value of $H_0$.

The implications of {\it M2} for the scalar spectral index $n_s$ are more interesting. As shown in figure~\ref{fig:AllPosteriors}, $n_s$ peaks in the range $0.935 \to 0.94$ in the {\it M2}-mode for all data combinations. 
Such a low spectral index is completely excluded in the standard $\Lambda$CDM model and, indeed, in most other extensions of $\Lambda$CDM we are aware of. From  the perspective of inflationary model building, $n_s \simeq 0.94$ could be realised, for example, by the Natural Inflation model \cite{Freese:1990rb,Freese:2014nla}, where such a low value of $n_s$ would allow the symmetry breaking scale to be sub-Planckian, thereby rendering radiative corrections to the potential subdominant (see, e.g., figure 88 of~\cite{Martin:2013tda}).

%%%%%%%%%%%%%%%%%%%%%%%%%%%%%%%%%%%%%%%%%%%%%%%%%%%%%%%%%
\section{Conclusions}
\label{Conclusions}

In this work, we have implemented into the  Boltzmann solver~{\sc class} the exact Boltzmann hierarchy that describes neutrino self-interaction via an effective four-fermion coupling first derived in~\cite{Oldengott:2014qra}, and used it to investigate the phenomenology of the CMB  in the presence of interacting neutrinos.
Along the way we have also compared our exact approach with two other models used in the literature: the ``separable ansatz'' or relaxation time approximation (RTA),  first introduced  in~\cite{Cyr-Racine:2013jua}, and the popular  $\left( c_{\text{eff}}^2,c_{\text{vis}}^2 \right)$-parameterisation.

While the agreement between our exact approach and the separable ansatz/RTA is remarkable {\it for this particular type of coupling}, both at the level of the neutrino fluid perturbations (i.e., density contrast, velocity divergence, etc.) and at the level of the CMB angular power spectrum, we caution that this is not a statement on the validity of the separable ansatz/RTA in general. A  self-interaction mediated by a massless scalar, for example, will most likely not respect the approximation, because the recoupling of neutrinos and especially their annihilation into scalars are expected to proceed in an energy-dependent fashion, causing the neutrino distribution to depart from a thermal shape for a period of time until the recoupling is complete.

The $\left( c_{\text{eff}}^2,c_{\text{vis}}^2 \right)$-parameterisation, on the other hand, is a poor model of neutrino scattering.  Previous works have already cast doubts on the physical meaningfulness of the model~\cite{Oldengott:2014qra,Cyr-Racine:2013jua,Sellentin:2014gaa}.  In this work, we have shown explicitly that even at the purely phenomenological level, the model fails to predict the correct scale dependence for the CMB temperature power spectrum.  Consequently, there is no meaningful way to map its model parameters to physical quantities such as the interaction strength.  We therefore strongly advocate against using this model as a phenomenological description of particle scattering for CMB anisotropy calculations.

Using the RTA we have furthermore derived constraints on the effective coupling constant $G_{\text{eff}}$ from cosmological observations in an MCMC analysis.
Interestingly, all data combinations used in the analysis yield a bimodal posterior distribution, wherein one mode represents the standard $\Lambda$CDM limit, and the other  a scenario in which neutrinos self-interact with an effective coupling constant $G_{\text{eff}} \simeq 0.03~{\rm MeV}^{-2} \simeq 3\times 10^9 \, G_{\text{F}}$. The latter, ``interacting'' mode is accompanied by an inferred scalar spectral index in the ballpark $n_{\text{s}} = 0.935 \to 0.94$, which may have interesting implications for inflationary model building.

\paragraph*{Note added:} While this work was in its final stages of completion, the preprint~\cite{Lancaster:2017ksf} appeared on arXiv, which likewise presented cosmological constraints on neutrino self-interactions described by a four-fermion coupling.  Although different methodologies have been applied, the results of~\cite{Lancaster:2017ksf} and our MCMC analysis are largely in agreement.

\section*{Acknowledgements}

We thank N.~Borghini, D.~Boriero, S.~Feld, D.~Grin, S.~Hannestad, D.~Schwarz, T.~Smith, and V.~Vennin for interesting discussions. 
IMO acknowledges the support by Studienstiftung des Deutschen Volkes and by RTG 1620 ``Models of Gravity'' funded by DFG.
TT acknowledges support from the Villum Foundation and computing resources from the Danish Center for Scientific Computing (DCSC).
The work of CR is supported by the DFG through the Transregional Research Center TRR33 ``The Dark Universe''.
The work of Y$^3$W is partially supported by the Australian Government through the Australian Research Council's Discovery Projects funding scheme (project DP170102382).

%%%%%%%%%%%%%%%%%%%%%%%%%%%%%%%%%%%%%%%%%%%%%%%%%%%%%%%%%

%\newpage
\appendix

\section{Proof of number-, energy- and momentum conservation}
\label{Proof of number-, energy- and momentum conservation}

We demonstrate analytically in this appendix that the Boltzmann hierarchy~\eqref{finalBoltzmann_mass} satisfies number, energy, and momentum conservation. In the case of the latter two, we furthermore show that our numerical implementation of~\eqref{finalBoltzmann_mass} respects these conservation laws.

We begin the analytical proof by first writing down the full expressions for the collision kernels $K^{\rm m}_{0}(q,q')$ and $K^{\rm m}_{1}(q,q')$  according to equation~\eqref{Kernel_Legendre}:
\begin{eqnarray}
\label{eq:k0}
K^{\rm m}_{0}(q,q') = && \frac{8}{q q'} \e^{
 \frac{1}{2} (q'-q)} \left[ \e^{-
    \frac{1}{2} (q + q')} \left(-q^2 (q'^2 +2 q' +2) - 
      2 q (q'^2 +3 q' +4 ) - 2 (q'^2 + 4 q' +8) \right) \right. \nonumber  \\
      && \left. + 2 \e^{-\frac{1}{2} |q - q'|} \left( 8 + q^2 - q q' + q'^2 + 4 |q - q'| \right) \right]\,, \\
K^{\rm m}_{1}(q,q') =&&  \frac{8}{q^2 q'^2} \e^{\frac{1}{2}(q'-q)} \left[ \e^{-\frac{1}{2}(q + q')} \left(12 q (q^2 + 9q + 42) + 
       2 q q' (5q^2 + 38q + 148) \right. \right.  \nonumber \\
      && \left. \left. + 4 q'^2 (q^3 + 6q^2 + 19 q+27) + q'^3 (2 + q)(q^2 + 2q + 6) + 504 (2 + q') \right) \right. \nonumber\\
  && \left. + 2 \e^{-\frac{1}{2} |q - q'|} \left( q^3 q' - 54 q'^2 
   - 54 q^2 - q^2 q'^2 + 104 q q' + q q'^3 - 504 \right) \right. \nonumber \\
    && \left. - 4 |q - q'| \e^{-\frac{1}{2} |q - q'|} \left(3 q^2 + 3 q'^2 - 5 q q' +126 \right) \right].    
\label{K0/K1}    
\end{eqnarray}
Conservation of number and energy density requires that the $\ell=0$ equation of~\eqref{finalBoltzmann_mass} vanishes under the appropriate momentum integration, while for momentum conservation the condition of a vanishing momentum integral applies to the $\ell=1$ equation.

\paragraph{Number conservation} 

Here we integrate the $\ell=0$ equation~\eqref{finalBoltzmann_mass}  by $\int \mathrm{d}q \, q^2 \bar{f} (q)$. This yields for the first collision term
\begin{equation}
-\frac{40}{3} \int \mathrm{d}q \, q^3 \bar{f} (q) \, \Psi_{\nu,0} (q).
\label{firstNumber} 
\end{equation}
Using  $K^{\rm m}_{0}(q,q')$ from equation~(\ref{eq:k0}), the same integration over momentum produces for the second collision term
\begin{equation}
\begin{aligned}
\int \mathrm{d} q' \, q'  \bar{f} (q') \, \Psi_{\nu,0} (q') \, \int \mathrm{d}q\, q \left[ K_0^{\text{m}}(q,q')-\frac{20}{9} q^2 \, q'^2 \e^{-q} \right] = \frac{40}{3} \int \mathrm{d}q' \, q'^3 \bar{f} (q') \, \Psi_{\nu,0} (q'),
\end{aligned}
\end{equation}
which exactly cancels the first term~\eqref{firstNumber}.

\paragraph{Energy conservation} Integrating the $\ell=0$ equation over $\int \mathrm{d}q \, q^3 \bar{f} (q)$ gives for the first collision term
\begin{equation}
-\frac{40}{3} \int \mathrm{d}q \, q^4 \bar{f} (q) \, \Psi_{\nu,0} (q),
\label{firstEnergy}
\end{equation}
which cancels out the second collision term, 
\begin{equation}
\begin{aligned}
\int \mathrm{d} q' \, q'  \bar{f} (q') \, \Psi_{\nu,0} (q') \, \int \mathrm{d}q\, q^2 \left[ K_0^{\text{m}}(q,q')+\frac{10}{9} q^2 \, q'^2 \e^{-q} \right] = \frac{40}{3} \int \mathrm{d}q' \, q'^4 \bar{f} (q') \, \Psi_{\nu,0} (q'),
\end{aligned}
\end{equation}
upon momentum-integration.

\paragraph{Momentum conservation} Integrating the $\ell=1$ equation over $\int \mathrm{d}q \, q^3 \bar{f} (q)$ leads to
\begin{equation}
-\frac{40}{3} \int \mathrm{d}q \, q^4 \bar{f} (q) \, \Psi_{\nu,1} (q),
\label{firstMomentum}
\end{equation}
which is exactly cancelled by the second collision term,
\begin{equation}
\begin{aligned}
\int \mathrm{d} q' \, q'  \bar{f} (q') \, \Psi_{\nu,0} (q') \, \int \mathrm{d}q\, q^2 \left[ K_1^{\text{m}}(q,q')-\frac{2}{9} q^2 \, q'^2 \e^{-q} \right] = \frac{40}{3} \int \mathrm{d}q' \, q'^4 \bar{f} (q') \, \Psi_{\nu,1} (q'),
\end{aligned}
\end{equation}
integrated over momentum in the same manner.

To demonstrate numerically that our implementation of~\eqref{finalBoltzmann_mass} in {\sc class} respects the conservation laws, we note first of all that  numerical cancellation of the collision terms is most challenging for large couplings~$G_{\text{eff}}$ and large wavenumbers~$k$.   In the former case, a large coupling causes the neutrino perturbations to undergo large-amplitude acoustic oscillations on sub-horizon scales, preserving power in the monopole and dipole.  In the latter case, the larger the wavenumber, the earlier it crosses the horizon and hence the longer time it spends in oscillations (and at a higher frequency).  See figures~\ref{TCAplot} and~\ref{perturbed}.  We therefore conclude that if numerical cancellation of the collision terms occurs for the largest relevant $G_{\rm eff}$ and $k$ values in the tightly-coupled limit, then our implementation is more than sufficiently accurate for all other situations.  

\begin{figure}[t]
\centering
\includegraphics[width=0.9\linewidth]{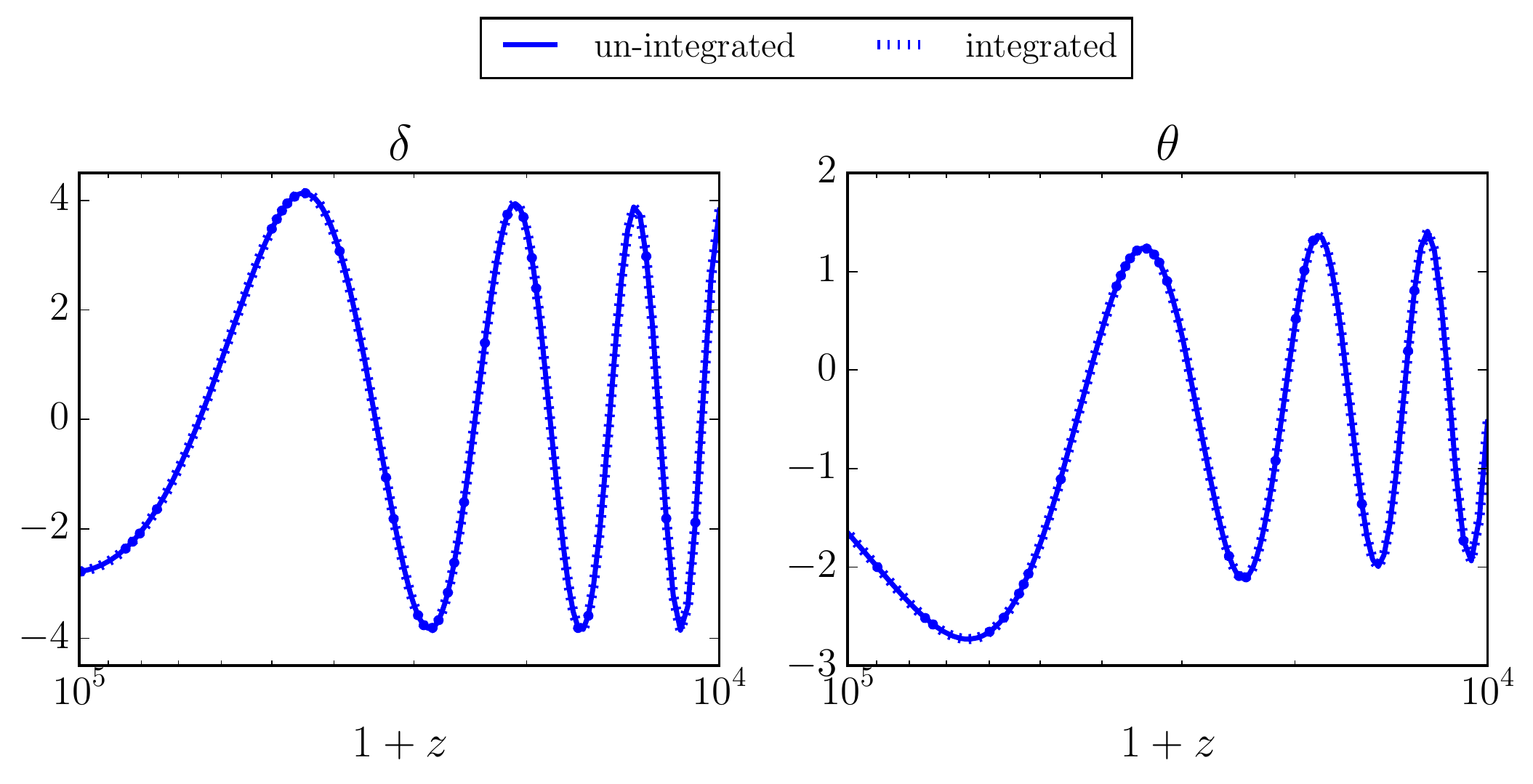}
\caption{Neutrino energy density contrast~$\delta$ (left) and velocity divergence~$\theta$ (right) as functions of $z$ in the tightly-coupled limit, with $k=1$~Mpc$^{-1}$. Blue solid lines denote solutions to the unintegrated equations~\eqref{TCA_unintegrated} assuming $G_{\text{eff}}=0.1$\,MeV$^{-2}$, whereas the dashed blue lines represent  solutions to the fluid equations~\eqref{TCA_equations}.}
\label{energy_momentum}
\end{figure}

In its unintegrated form, the tightly-coupled limit of the Boltzmann hierarchy reads
\begin{equation}
\begin{aligned} 
 \dot{\Psi}_{0}(q) =&  - k \Psi_{1}(q) + \frac{1}{6} \frac{\partial \ln \bar{f}}{\partial \ln q} \dot{h} - \frac{40}{3} \frac{2\mathrm{N} T_{\nu,0}^5 G_{\text{eff}}^2}{a^4(2\pi)^{3}} \, q \,
 \Psi_{0}(q)  \\
 &\hspace{10mm}+ \frac{2\mathrm{N} T_{\nu,0}^5 G_{\text{eff}}^2}{a^4(2\pi)^{3}} \int  \dd q' \left[ 2 K^{\rm m}_0(q,q') - \frac{20}{9} q^2\, {q'}^2 \e^{-q}
  \right] \,\frac{q' \bar{f}(q')}{q \bar{f}(q)}   \, \Psi_{0}(q') \, , \\ 
\dot{\Psi}_{1}(q) =&  \frac{1}{3}k \Psi_{0}(q) - \frac{40}{3} \frac{2\mathrm{N} T_{\nu,0}^5 G_{\text{eff}}^2}{a^4(2\pi)^{3}} \, q \, \Psi_{1}(q) \\
&\hspace{10mm}+ \frac{2\mathrm{N} T_{\nu,0}^5 G_{\text{eff}}^2}{a^4(2\pi)^{3}} \int  \dd q'  \left[ 2 K^{\rm m}_1(q,q')
+\frac{10}{9} q^2 \, {q'}^2 \e^{-q} \right] \, \frac{q' \bar{f}(q')}{q \bar{f}(q)}  \, \Psi_{1}(q') \, ,
\end{aligned}
\label{TCA_unintegrated}
\end{equation}  
where upon integration in momentum we should recover the fluid equations~\eqref{TCA_equations}.   Thus, the solutions to equations~\eqref{TCA_unintegrated} and~\eqref{TCA_equations} must yield the same $\delta(\eta)$ and $\theta(\eta)$ if energy and momentum conservation are respected numerically by our implementation.   As shown in figure~\ref{energy_momentum},  in which we plot the evolution of $\delta$ and $\theta$ computed from both equations~\eqref{TCA_unintegrated} and~\eqref{TCA_equations} for $k=1$~Mpc$^{-1}$ and $G_{\text{eff}}=0.1$\,MeV$^{-2}$, this is indeed the case.

%%%%%%%%%%%%%%%%%%%%%%%%%
%%%%%%%%%%%%%%%%%%%%%%%%%

\bibliographystyle{utcaps}
\bibliography{Literature}

%%%%%%%%%%%%%%%%%%%%%%%%%%%%%%%%%%%%%%%%%%%

%%%%%%%%%%%%%%%%%%%%%%%%%%%%%%%%%%%%%%%%%%%%%%%%%%%%%%%%%%%%%%%%%%%%%%%%%%%%%%%%%%%%%

\end{document}